\documentclass[apj] {emulateapj}
\usepackage{apjfonts}
\usepackage{epstopdf}
\usepackage{amsmath}
\usepackage{subfigure}

\def\wig#1{\mathrel{\hbox{\hbox to 0pt{%
          \lower.5ex\hbox{$\sim$}\hss}\raise.4ex\hbox{$#1$}}}}

\shorttitle{Super Earths with Flat Transmission Spectra}

\newcommand{\mj}{$M_{\mathrm{J}}$}

\newcommand{\me}{$M_{\oplus}$}
\newcommand{\re}{$R_{\oplus}$}

\newcommand{\teff}{$T_{\rm eff}$}

\newcommand{\cp}{\citep}
\newcommand{\ct}{\citet}

\newcommand{\kepler}{\emph{Kepler}}

\newcommand{\icarus}{Icarus} 
\newcommand{\fsed}{$f_{\rm sed}$} 
\newcommand{\fhaze}{$f_{\rm haze}$} 
\newcommand{\nas}{Na$_2$S}

\newcommand{\kzz}{$K_{zz}$}
\newcommand{\jwst}{\emph{JWST}}

\slugcomment{Draft for ApJ}

\begin{document}

\title{Thermal Emission and Albedo Spectra of Super Earths with Flat Transmission Spectra}

\author{Caroline V. Morley\altaffilmark{1}, Jonathan J. Fortney\altaffilmark{1}, Mark S. Marley\altaffilmark{2}, Kevin Zahnle\altaffilmark{2}, Michael Line\altaffilmark{2}, Eliza Kempton\altaffilmark{3}, Nikole Lewis\altaffilmark{4}, Kerri Cahoy\altaffilmark{5}}

\altaffiltext{1}{Department of Astronomy and Astrophysics, University of California, Santa Cruz, CA 95064; cmorley@ucolick.org}
\altaffiltext{2}{NASA Ames Research Center} 
\altaffiltext{3}{Grinnell College} 
\altaffiltext{4}{Space Telescope Science Institute} 
\altaffiltext{5}{Massachusetts Institute of Technology} 

\begin{abstract}
 
 Planets larger than Earth and smaller than Neptune are some of the most numerous in the galaxy, but observational efforts to understand this population have proved challenging because optically thick clouds or hazes at high altitudes obscure molecular features \cp{Kreidberg14}. We present models of super Earths that include thick clouds and hazes and predict their transmission, thermal emission, and reflected light spectra. Very thick, lofted clouds of salts or sulfides in high metallicity (1000$\times$ solar) atmospheres create featureless transmission spectra in the near-infrared. Photochemical hazes with a range of particle sizes also create featureless transmission spectra at lower metallicities. Cloudy thermal emission spectra have muted features more like blackbodies, and hazy thermal emission spectra have emission features caused by an inversion layer at altitudes where the haze forms. Close analysis of reflected light from warm ($\sim$400-800 K) planets can distinguish cloudy spectra, which have moderate albedos (0.05--0.20), from hazy models, which are very dark (0.0--0.03). Reflected light spectra of cold planets ($\sim$200 K) accessible to a space-based visible light coronagraph will have high albedos and large molecular features that will allow them to be more easily characterized than the warmer transiting planets. We suggest a number of complementary observations to characterize this population of planets, including transmission spectra of hot ($\gtrsim1000$ K) targets, thermal emission spectra of warm targets using the James Webb Space Telescope (\jwst), high spectral resolution (R$\sim$10$^5$) observations of cloudy targets, and reflected light spectral observations of directly-imaged cold targets. Despite the dearth of features observed in super Earth transmission spectra to date, different observations will provide rich diagnostics of their atmospheres.  

\end{abstract}

 
\section{Introduction}

Since its launch in 2008, the \emph{Kepler} mission has revealed a population of planets with radii between that of Earth and Neptune, which make up a substantial fraction of the planets in the galaxy \cp{Borucki11, Howard12}. No planet of that size exists in our own solar system as an archetype for these ``super Earths.''\footnote{We use the term `super Earth' here to mean planets larger than Earth and smaller than Neptune, but recognize these planets are diverse in their compositions and many may be more accurately considered `sub Neptunes.'} This population likely has a range of compositions from rocky, to water-rich, to gas-rich \cp{Rogers15,Wolfgang14}, but we have not yet probed their compositions directly. A critical part of the puzzle to understand the nature of super Earths is to measure the abundances of molecules in their atmospheres. 

One powerful tool that has been used to probe the atmospheres of transiting planets is transmission spectroscopy. During a transit, the transit depth is measured simultaneously at multiple wavelengths. At wavelengths of strong absorption features, the planet's atmosphere will become optically thick at a higher altitude and we will observe a deeper transit; at wavelengths outside these absorption features, the planet's atmosphere is optically thinner and the transit depth is shallower. The depth of the features we observe in a cloud-free atmosphere scales linearly with the pressure scale height $H$ \cp{Seager00, Hubbard01}. The scale height is defined as $H=kT/\mu g$, where $k$ is Boltzmann's constant, $T$ is temperature, $g$ is gravity, and $\mu$ is the mean molecular weight. In the absence of clouds, low gravity, low density targets have the largest amplitude features and many have been targeted for characterization. 

By measuring the amplitude of features in a planet's transmission spectrum we can both probe the composition of absorbers like sodium, potassium, methane, water, and carbon monoxide and also measure the bulk composition of the atmosphere by measuring the mean molecular weight. If the observed mean molecular weight is low ($\mu\sim2.3$) the planet is H/He-rich like a scaled down Neptune; if it is high, it may be water, nitrogen, or carbon dioxide rich, more akin to a terrestrial planet \cp{Miller-Ricci09}. 

\subsection{Observations of Super Earths}

Hundreds of orbits of Hubble Space Telescope (\emph{HST}) time have been dedicated to characterizing these small planets, as well as hundreds of hours of ground-based observations. Despite this dedication of resources, super Earths and sub-Neptunes have proved extremely challenging to characterize with this technique because their features are more muted than predicted using cloud-free models. 

By far the most-studied super Earth to date is GJ 1214b, the first planet discovered by the MEarth survey \cp{Charbonneau09}. GJ 1214b is a 6.16 $\pm$0.91\me\ and 2.71$\pm$0.24\re\ planet, and, critically, orbits a fairly bright mid M dwarf (M4.5). Its transit depth of over 1\% and a short orbital period of 38 hours make it an ideal target for high signal-to-noise followup observations. 

Early observations from both ground and space were inconclusive: they showed no features, but were not sensitive enough to detect the small features predicted for a high mean molecular weight atmosphere \cp{Bean10, Desert11c, Crossfield11, Croll11, Berta12, deMooij12, Murgas12, Teske13, Fraine13}. In 2014, \ct{Kreidberg14} measured 15 additional transits of GJ 1214b with \emph{HST} Wide Field Camera 3 (WFC3) grism spectroscopy (1.1--1.7 $\mu$m) and detected, at high signal to noise, a featureless transmission spectrum. Unlike the previous observations, these observations were sensitive enough to detect features in a high mean molecular weight atmosphere. They concluded that the predicted molecular features are obscured by a high altitude cloud or haze layer. 

Other planets close to GJ 1214b's size have also been observed with this technique, with somewhat lower signal-to-noise than the \ct{Kreidberg14} observations. \ct{Knutson14b} present observations of the super Earth HD 97658b and show that its spectrum is consistent with a flat line. Likewise, the Neptune-sized GJ 436b and GJ 3470b also have featureless spectra measured with WFC3 within their measurement uncertainties \cp{Knutson14a, Ehrenreich14}. In fact, the only planet in the super-Earth to Neptune mass range with a statistically significant spectral feature is HAT-P-11b; water vapor absorption was detected using WFC3 with an amplitude of 250 parts per million \cp{Fraine14}. This measurement is consistent with a metal-enhanced H/He dominated atmosphere with a several hundred times solar metallicity composition or a less enriched atmosphere with features muted by clouds or hazes. 

The observing efforts to date have revealed that small, cool planets have relatively featureless transmission spectra. If features are muted in the transmission spectra of all small planets, it will be extremely challenging to characterize their compositions using transmission spectroscopy. 

\subsection{Understanding Super Earths Despite the Clouds}

While these featureless near-infrared transmission spectra are informative---they inform us that there is an optically thick, gray absorber in the measured wavelength range---they do not allow us to measure the composition of the atmosphere. To understand the compositions of super Earths---perhaps the most abundant planets in the galaxy---we need to probe their atmospheres with other techniques. A number of pathways will help to accomplish this goal, including transmission spectra of hotter targets, thermal emission spectra, and reflected light spectra.

In this paper, we use models of super Earths to understand how we can characterize super Earths as a class. We move beyond modeling GJ 1214b itself and run models of its cousins, with the same gravity and host star but different incident flux. Cloud and haze formation depends strongly on incident flux (and resulting equilibrium temperature) so spanning a range of irradiation levels allows us to make predictions about a diverse set of planets. 

We quantify properties of clouds or hazes thick enough to flatten transmission spectra at the signal-to-noise of the \ct{Kreidberg14} observations for a variety of different incident flux levels. Using these cloud properties, we generate both thermal emission spectra and reflected light spectra. With the upcoming \jwst\ mission, thermal emission of selected super Earths will be observable over a wide wavelength range \cp{Gardner06}; we show how optically thick clouds and hazes will shape that thermal emission. In the more distant future, missions to detect reflected light from exoplanets using a space-based coronagraph are being planned \cp{Spergel15}. We show that reflected light spectra will be a promising technique to understand very cloudy super Earths, especially for colder objects.

\subsection{Format of this Paper}

In Section \ref{methods}, we describe the extensive set of modeling tools used to model the effects of clouds and hazes on super Earth spectra. In Section \ref{eq_clouds}, transmission, emission, and reflection spectra for planets with equilibrium clouds (both salt/sulfide in warm planets and water ice in cold planets) are presented. In Section \ref{phot_hazes}, transmission, emission, and reflection spectra for planets with photochemical hazes are presented. In Section \ref{discussion}, we discuss implications of this work for future studies and in Section \ref{conclusion} we conclude. The Appendix discusses the new, flexible radiative transfer tool developed for this work. 

\section{Methods} \label{methods}
 
In order to predict spectra of small planets with clouds and hazes, we use a comprehensive suite of atmosphere modeling tools. We use a 1D radiative--convective model to calculate the pressure--temperature structure, a photochemical model to calculate the formation of soot precursors (hydrocarbons that may form hazes), and a cloud model to calculate cloud altitudes, mixing ratios, and particle sizes. We then calculate spectra in different geometries and wavelengths using a transmission spectrum model, a thermal emission spectrum model, and an albedo model. In the following subsections we discuss each of these calculations. 

We run a grid of radiative-convective models of GJ 1214b analogs (g=7.65 m/s$^2$, M4.5 host star). We vary the distance from the host star to encapsulate a range of super Earths from temperatures of 190--1400 K (0.01--30$\times$ GJ 1214b's incident flux). In one set of models, we include ``equilibrium clouds''. These, in this work, are considered to be clouds that form when the pressure of a condensible gas exceeds the saturation vapor pressure; we assume that all material in excess of the saturation vapor pressure condenses into cloud material. For these objects, the clouds include water ice (for the coldest models), and salts and sulfides (for the warmer models). In the other set of models, we include a photochemical haze using a photochemical model.

Given the number steps involved for each set of models, we will first outline the modeling process performed for every set of parameters. We follow a slightly different set of steps for the equilibrium clouds and the photochemical hazes. 

\paragraph{Equilibrium clouds}
\begin{enumerate} \itemsep1pt \parskip0pt \parsep0pt

\item Generate a cloud-free pressure--temperature (P--T) profile at high metallicity (100--1000$\times$ solar metallicity) using a modified 1D radiative--convective model \label{radconitem}
\item Using that P--T profile (1), calculate cloud locations, particle sizes, and optical properties using a stand-alone version of the \ct{AM01} cloud code
\item Using that P--T profile (1), calculate the equilibrium chemistry along the profile using Chemical Equilibrium with Applications (CEA). 
\item Using the P--T profile, cloud output, and equilibrium chemistry (1,2,3), calculate the model transmission spectrum and compare it to the flat \cp{Kreidberg14} spectrum of GJ 1214b. 
\item Using the P--T profile, cloud output, and equilibrium chemistry (1,2,3), calculate the thermal emission spectrum 
\item Using the P--T profile, cloud output, and equilibrium chemistry (1,2,3), calculate the reflected light spectrum

\end{enumerate}

\paragraph{Photochemical hazes}

\begin{enumerate} \itemsep1pt \parskip0pt \parsep0pt

\item Using a pre-computed pressure--temperature profile, calculate the disequilibrium chemistry caused by vertical mixing and photochemistry
\item Using the abundances and locations of soot precursors from (1), calculate a pressure--temperature profile consistent with haze using a 1D radiative--convective atmosphere model
\item Using the P--T profile and haze properties (2), calculate the model transmission spectrum and compare to the flat \cp{Kreidberg14} spectrum of GJ 1214b. 
\item Using the P--T profile and haze properties (2), calculate the thermal emission spectrum 
\item Using the P--T profile and haze properties (2), calculate the reflected light spectrum

\end{enumerate}

\subsection{1D Radiative--Convective Model}

For objects with and without clouds, we calculate their temperature structures assuming 1D atmospheres in radiative--convective equilibrium. Our approach has been successfully applied to objects ranging in size from moons to brown dwarfs; the models are described in \ct{Mckay89, Marley96, Burrows97, MM99, Marley02, Fortney05, Saumon08, Fortney08b}. 

We use the radiative transfer techniques described in \ct{Toon89} and use Mie theory to calculate the absorption and scattering of cloud particles in each layer of the atmosphere. The opacity database for gases is described extensively in \ct{Freedman08}. In this work, the opacity database includes two significant updates since \ct{Freedman08}, which are described in \ct{Saumon12}: a new molecular line list for ammonia \cp{Yurchenko11} and an improved treatment of collision induced H$_2$ absorption \cp{Richard12}. Optical properties for salts and sulfides are as described in \ct{Morley12}; for ZnS and KCl they are obtained from \ct{Querry87} and for Na$_2$S we combine laboratory and numerical measurements from \ct{Montaner79} and \ct{Khachai09}. 

The opacities, using the k-coefficient technique for computational speed and accuracy, are pre-calculated and pre-summed at multiples of solar metallicity ranging from [M/H]=0.0 to 1.7 (1--50$\times$ solar), but super Earths potentially have much higher metallicity atmospheres (see \ct{Fortney13} and discussion in Section \ref{highmetlikely}). Higher metallicity opacities have not been calculated, so in order to calculate the temperature structures at higher metallicities (100-1000$\times$ solar), we approximate the gas opacity by multiplying the [M/H]=1.7 pre-summed opacities by the appropriate factor. For example, for 300$\times$ solar metallicity, we multiply the 50$\times$ solar summed molecular gas opacities by 6. We decrease the abundance of hydrogen and helium by the same proportion and calculate the collision induced absorption separately from the other molecular gas opacities. This approximation is appropriate for the qualitative results explored here; for future work, e.g. comparing models to data, new k-coefficients at 100--1000$\times$ solar metallicity should be used. 

\begin{figure}[t]
  \center   \includegraphics[width=3.5in]{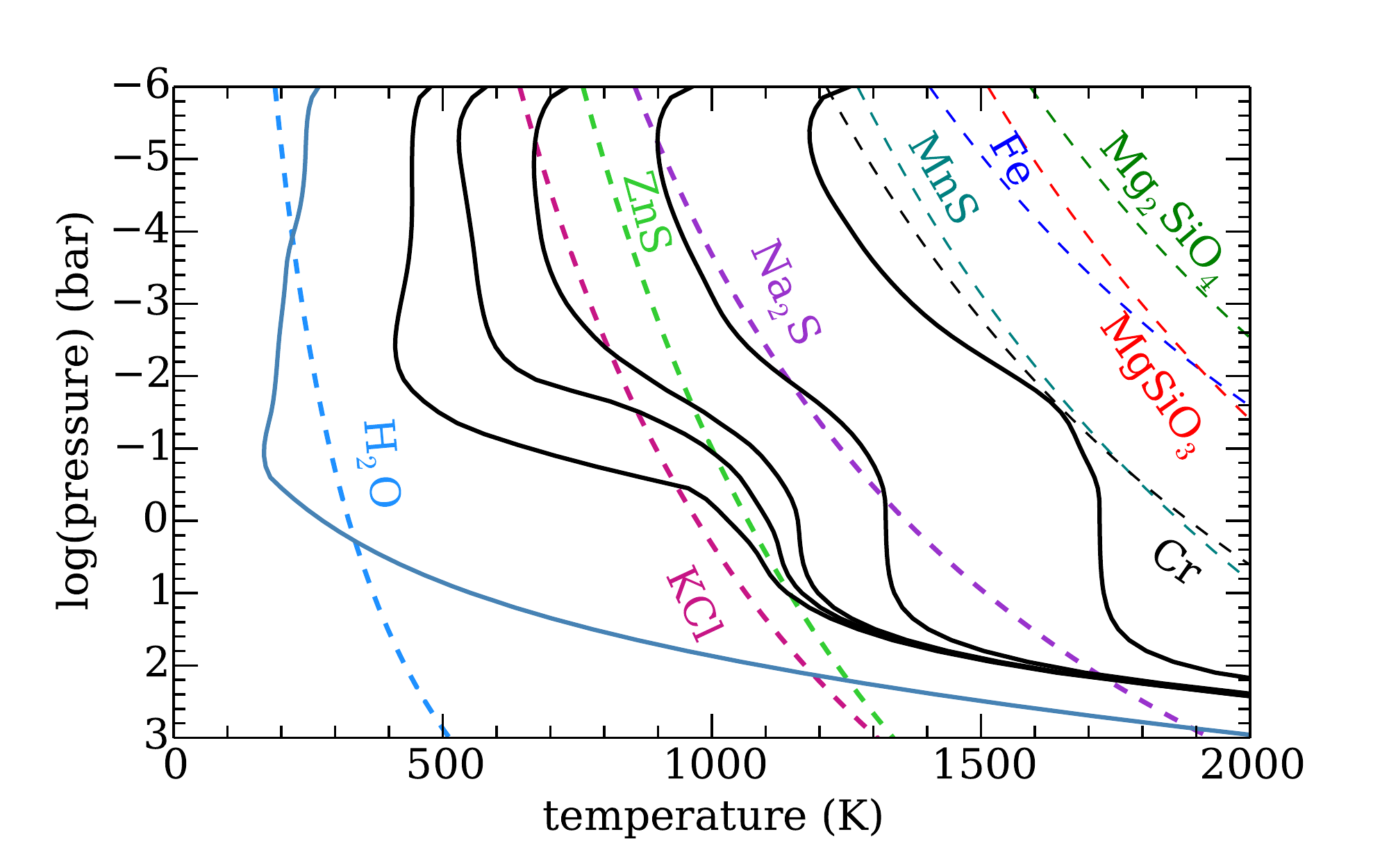}
 \caption{Pressure--temperature profiles of models at 300$\times$ solar metallicity with cloud condensation curves. P--T profiles are shown as solid curves; black indicates models with salt/sulfide clouds and blue indicates models with water ice clouds. From left to right, these profiles are at 0.01, 0.3, 1, 3, 10, and 30 $\times$ GJ 1214b's incident flux. Condensation curves are shown as dashed lines for individual cloud species; a cloud forms where the P--T profile crosses the condensation curve.   }
\label{ptprofs_eq}
\end{figure}

Examples of calculated P--T profiles are shown in Figure \ref{ptprofs_eq}, for models from 0.01 to 30$\times$ GJ 1214b's incident flux. 

\subsection{Equilibrium Chemistry}

After we calculate the pressure--temperature profiles of models with greater than 50$\times$ solar metallicity, we calculate the composition, assuming chemical equilibrium, along that profile. To be clear, this isn't strictly self-consistent. However, tests run using 10$\times$ solar and 50$\times$ solar compositions---both of which we calculate self-consistently with the chemistry---show that the effect on emergent spectra is very small. 

For this calculation, we use the Chemical Equilibrium with Applications model (CEA, Gordon \& McBride 1994) to compute the thermochemical equilibrium molecular mixing ratios (with applications to exoplanets see, \ct{Visscher10, Line10, Moses11, Line11, Line13}).  CEA minimizes the Gibbs Free Energy with an elemental mass balance constraint of a parcel of gas given a local temperature, pressure, and elemental abundances. We include species that contain H, C, O, N, S, P, He, Fe, Ti, V, Na, and K. We account for the depletion of oxygen due to enstatite condensation by removing 3.28 oxygen atoms per Si atom \cp{Burrows99}.   When adjusting the metallicity all relative elemental abundances are rescaled equally relative to H while ensuring that the elemental abundances sum to one.  

\subsection{Cloud Model}

 \begin{figure}[ht]
  \center   \includegraphics[width=3.4in]{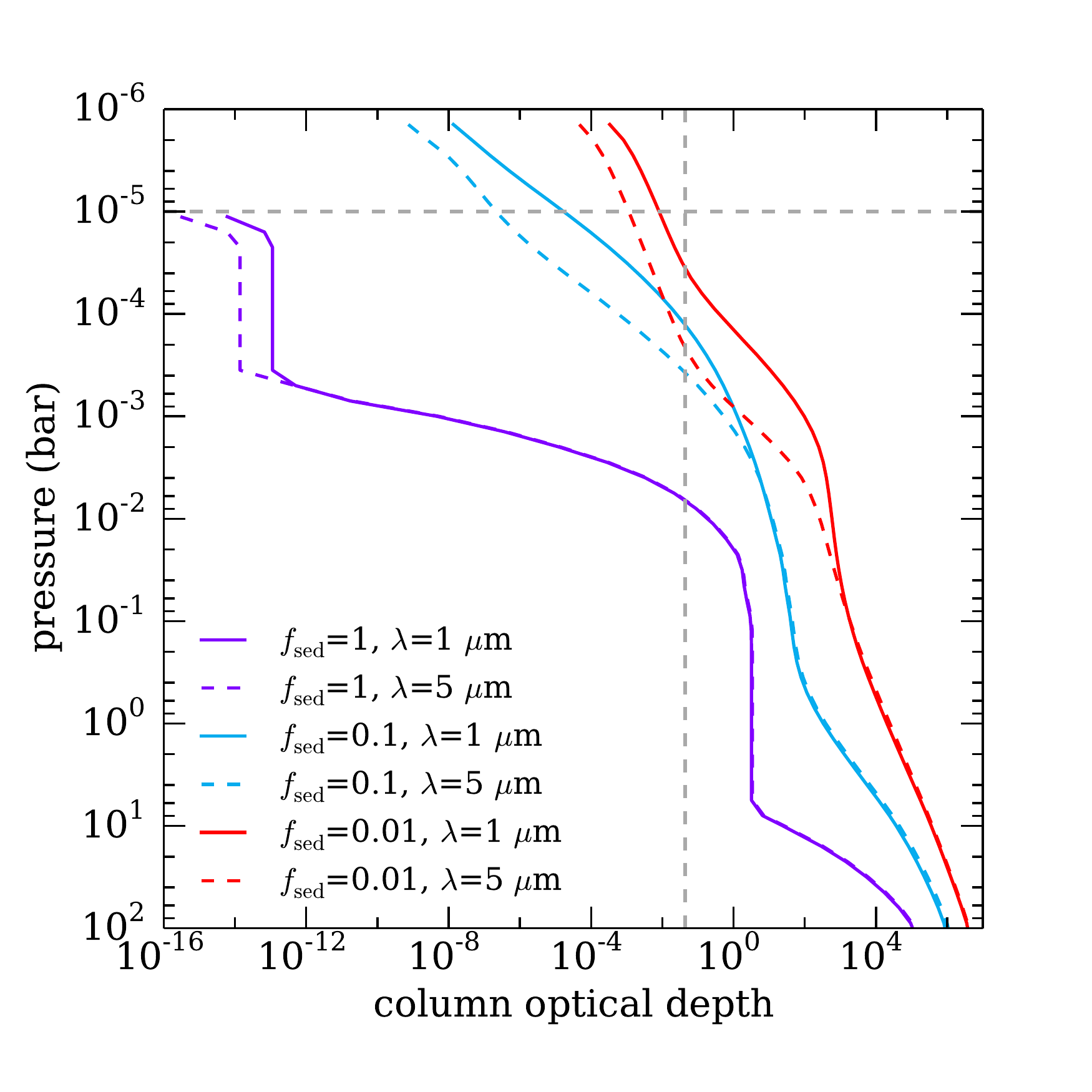}
	\vspace{-1.4cm}
  \center   \includegraphics[width=3.4in]{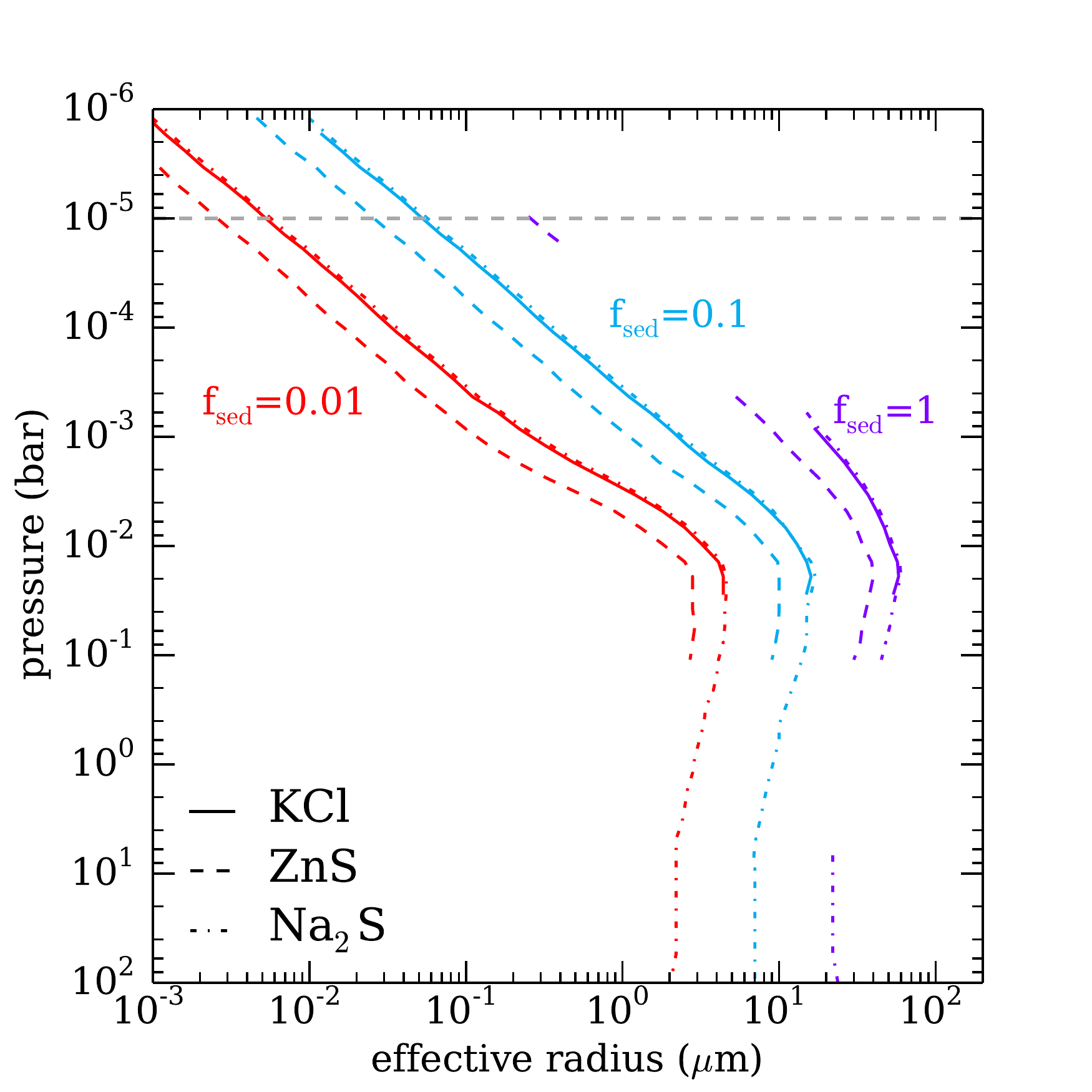}
 \caption{Column optical depth and mode particle sizes of clouds with varied sedimentation efficiency \fsed, 300$\times$ solar metallicity composition, and GJ 1214b's incident flux. Top panel shows the column optical depth at two wavelengths (1 and 5 \micron) as a function of pressure for \nas, KCl, and ZnS clouds (summed), with \fsed\ from 0.01 to 1. Note that lower \fsed\ values result in optically thicker clouds at higher altitudes. The dashed vertical gray line shows the $\tau=1$ line for slant viewing geometry using equation 6 from \ct{Fortney05c}. The bottom panel shows the mode particle size of each cloud species for 3 values of \fsed; note that lower \fsed\ values result in very small particles. The dashed horizontal gray line in both panels shows the approximate altitude of GJ 1214b's cloud to cause a flat transmission spectrum. }
\label{cloudcode}
\end{figure}

 \begin{figure}[ht]
  \center   \includegraphics[width=3.4in]{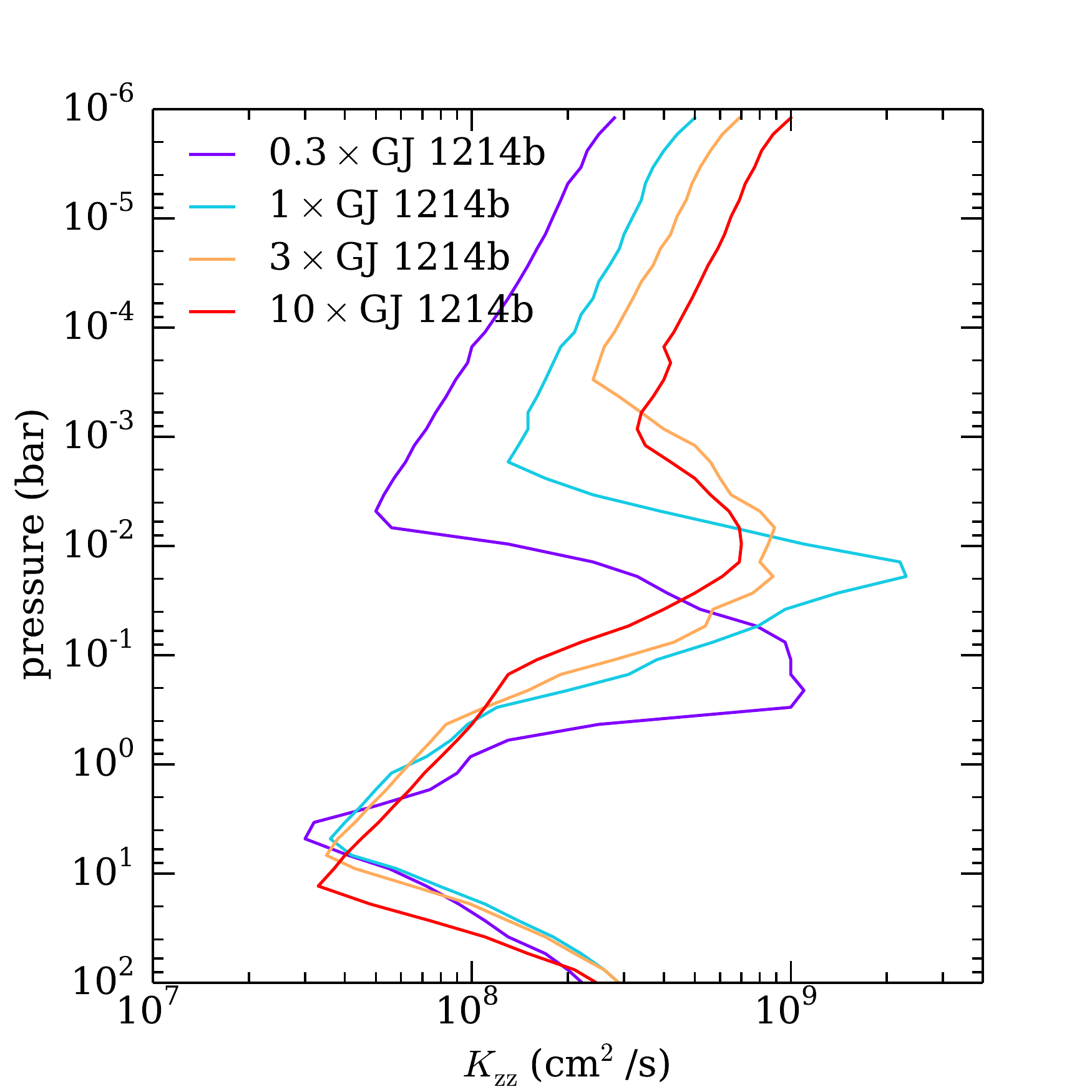}
 \caption{Eddy diffusion coefficients (\kzz) calculated within the \ct{AM01} cloud code for models with 300$\times$ solar composition and 0.3--10$\times$ the incident flux of GJ 1214b.}
\label{cloudcode_kzz}
\end{figure}

We use a modified version of the \ct{AM01} cloud model which includes sulfide and salt clouds \cp{Morley12, Morley13}. 
The \ct{AM01} approach balances the upward transport of vapor and condensate by turbulent mixing in the atmosphere with the downward transport of condensate by sedimentation using the equation 

\begin{equation} \label{advdiff}
-K_{zz} \frac{\partial q_t}{ \partial z} - f_{\textrm{sed}} w_*q_c =0, 
\end{equation}
where $K_{zz}$ is the vertical eddy diffusion coefficient, $q_t$ is the mixing ratio of condensate and vapor, $q_c$ is the mixing ratio of condensate, $w_*$ is the convective velocity scale, and \fsed\ is a parameter that describes the efficiency of sedimentation in the atmosphere. \fsed\ is the only tunable free parameter in this cloud mode. It represents the ratio of the sedimentation velocity to the convective velocity. Higher \fsed\ values result in larger particles in a vertically compact layer; lower \fsed\ values result in smaller particles in a more lofted cloud layer. Typical \fsed\ values for brown dwarfs, for which this model was first developed, are 1--5 \cp{Saumon08, Morley12}, while planets may have clouds best fit with smaller \fsed\ (<1) \cp{AM01, Morley13}. 

Cloud material in excess of the saturation vapor pressure of the limiting gas is assumed to condense into cloud particles. We extrapolate the saturation vapor pressure equations from \ct{Morley12} to high metallicites, which introduces some uncertainties but serves as a reasonable first-order approximation for the formation of these cloud species. 

We prescribe a lognormal size distribution of particles given by
\begin{equation}
\dfrac{dn}{dr}=\dfrac{N}{r\sqrt{2\pi}\ln\sigma}\exp{\left[-\dfrac{\ln^2(r/r_g)} {2\ln^2\sigma}\right]}
\end{equation}
where $N$ is the total number concentration of particles, $r_g$ is the geometric mean radius, and $\sigma$ is the geometric standard deviation. $\sigma$ is fixed (2.0) for this study and falling speeds of particles within this distribution are calculated assuming viscous flow around spheres (and using the Cunningham slip factor to account for gas kinetic effects). We calculate the other parameters in equation \ref{advdiff} ($K_{zz}$ and $w_*$) using mixing length theory to relate turbulent mixing to the convective heat flow \cp{Gierasch85}. Rigorously the convective heat flow becomes zero well above the radiative-convective boundary. However for purposes only of computing \kzz\ we impose a very small convective heat flux through the radiative stratosphere, causing \kzz\ to increase with altitude at the top of the atmosphere. A lower bound for $K_{zz}$ of 10$^5$ cm$^2$s$^{-1}$ represents the residual turbulence from processes such as breaking gravity waves in radiative regions. \kzz values for representative models are shown in Figure \ref{cloudcode_kzz}; the values we calculate qualitatively match the values found by recent 3D modeling efforts \cp[their Figure 13]{Charnay15}, and are generally between $10^8$ and $10^9$ cm$^2$s$^{-1}$ in the upper atmosphere. The good agreement with the \ct{Charnay15} \kzz\ profiles validates our approach.

Examples of the calculated cloud properties (cloud optical depth and particle size) are shown in Figure \ref{cloudcode} as a function of a free parameter in this prescription, \fsed. The top panel of Figure \ref{cloudcode} shows the resulting column optical depth of the cloud material at $\lambda$=1 and 5 \micron. Note that the only \fsed\ value shown that results in optically thick clouds at high altitude is \fsed=0.01. The lower panel of Figure \ref{cloudcode} shows the particle sizes for each cloud for three different \fsed\ values. \fsed=0.01 results in very small particles (0.01--0.1 \micron) at the cloud top; larger \fsed\ values result in larger particles (0.1--100 \micron). 

Two versions of the \ct{AM01} code are frequently used. One version is coupled self-consistently to the calculation of radiative--convective equilibrium; the other is a stand-alone version which calculates the clouds along a given P--T profile without recalculating the profile self-consistently. Note that the convective heat flow for a cloud-free model is used in the calculation of \kzz\ in the stand-alone version. Here we use the uncoupled, stand-alone version for higher metallicity calculations (100--1000$\times$ solar) for which the convergence for self-consistent models is numerically challenging. The pressure--temperature profiles for the models with photochemical haze are calculated self-consistently with the opacity of the hazes, but the haze properties are not calculated within the \ct{AM01} framework.

\subsection{Photochemistry}

 \begin{figure}[t]
  \center   \includegraphics[width=3.4in]{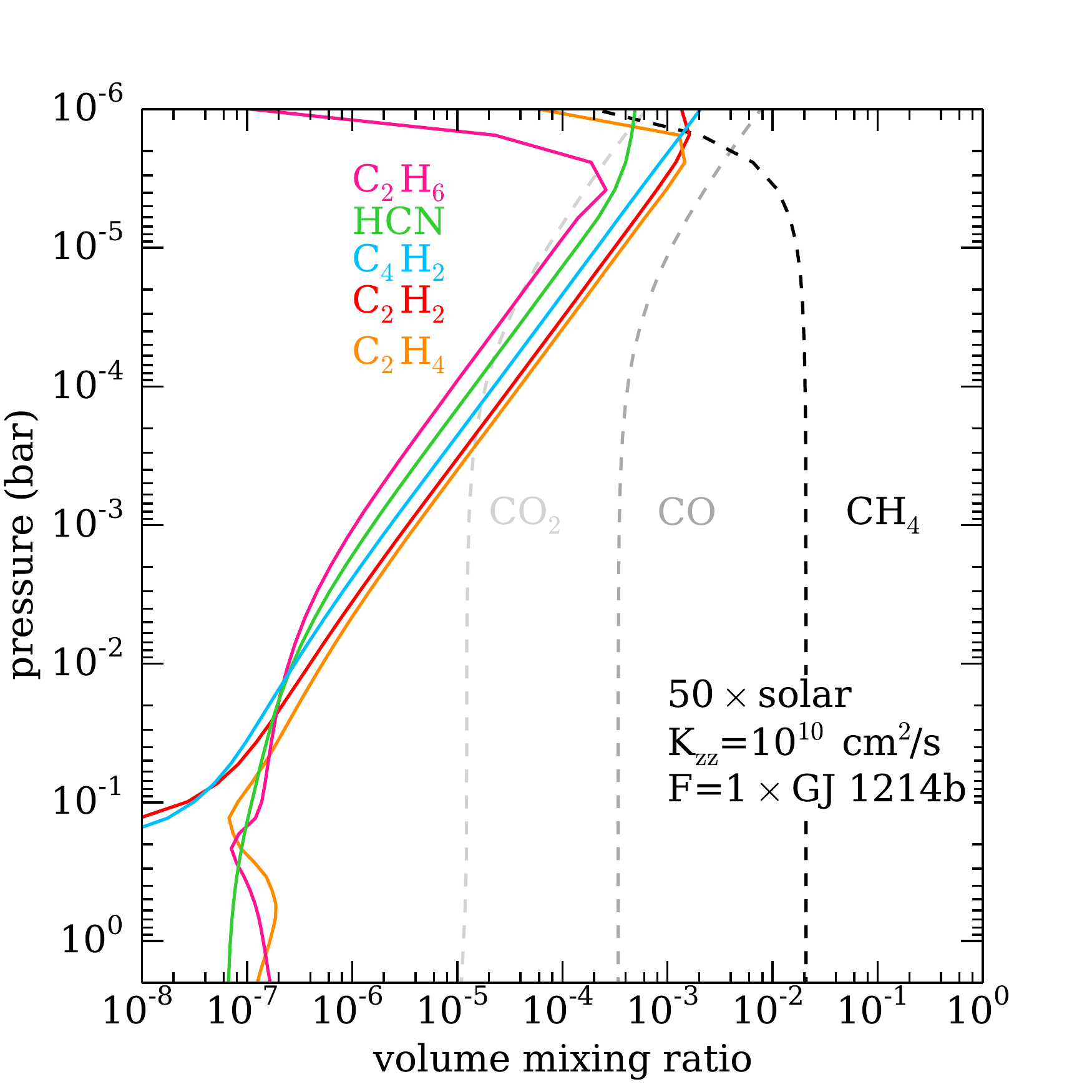}
 \caption{Carbon photochemistry for a 50$\times$ solar metallicity model with GJ 1214b's incident flux and \kzz=10$^{10}$ cm$^2$ s$^{-1}$. Soot precursors (solid lines) like C$_2$H$_2$, C$_2$H$_4$, C$_2$H$_6$, C$_4$H$_2$, and HCN form in the upper layers of the atmosphere where methane is dissociated by UV flux from the star. Other major carbon-bearing species are shown as dashed lines. }
\label{photochem_single}
\end{figure}

 \begin{figure}[t]
  \center   \includegraphics[width=3.4in]{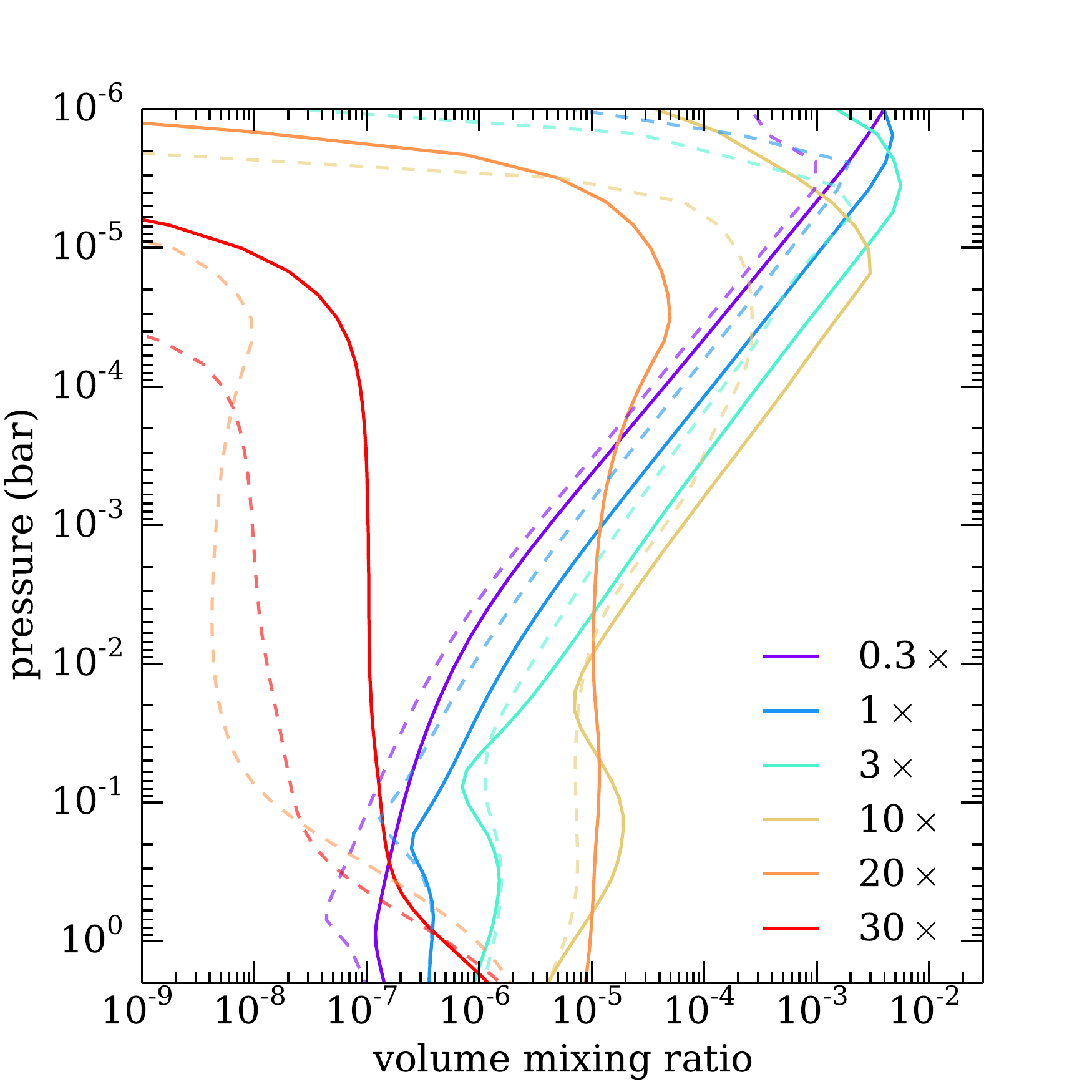}
 \caption{Carbon photochemistry for a set of 50$\times$ solar metallicity models with varied incident flux. Lines show sum of mixing ratios of all soot precursors. Solid lines show \kzz=10$^{10}$ cm$^2$ s$^{-1}$; dashed lines show \kzz=10$^8$ cm$^2$ s$^{-1}$. Note that soot precursor production peaks at 1--3$\times$ the irradiation of GJ 1214b. }
\label{photochem_summed}
\end{figure}

 \begin{figure}[t]
  \center   \includegraphics[width=3.4in]{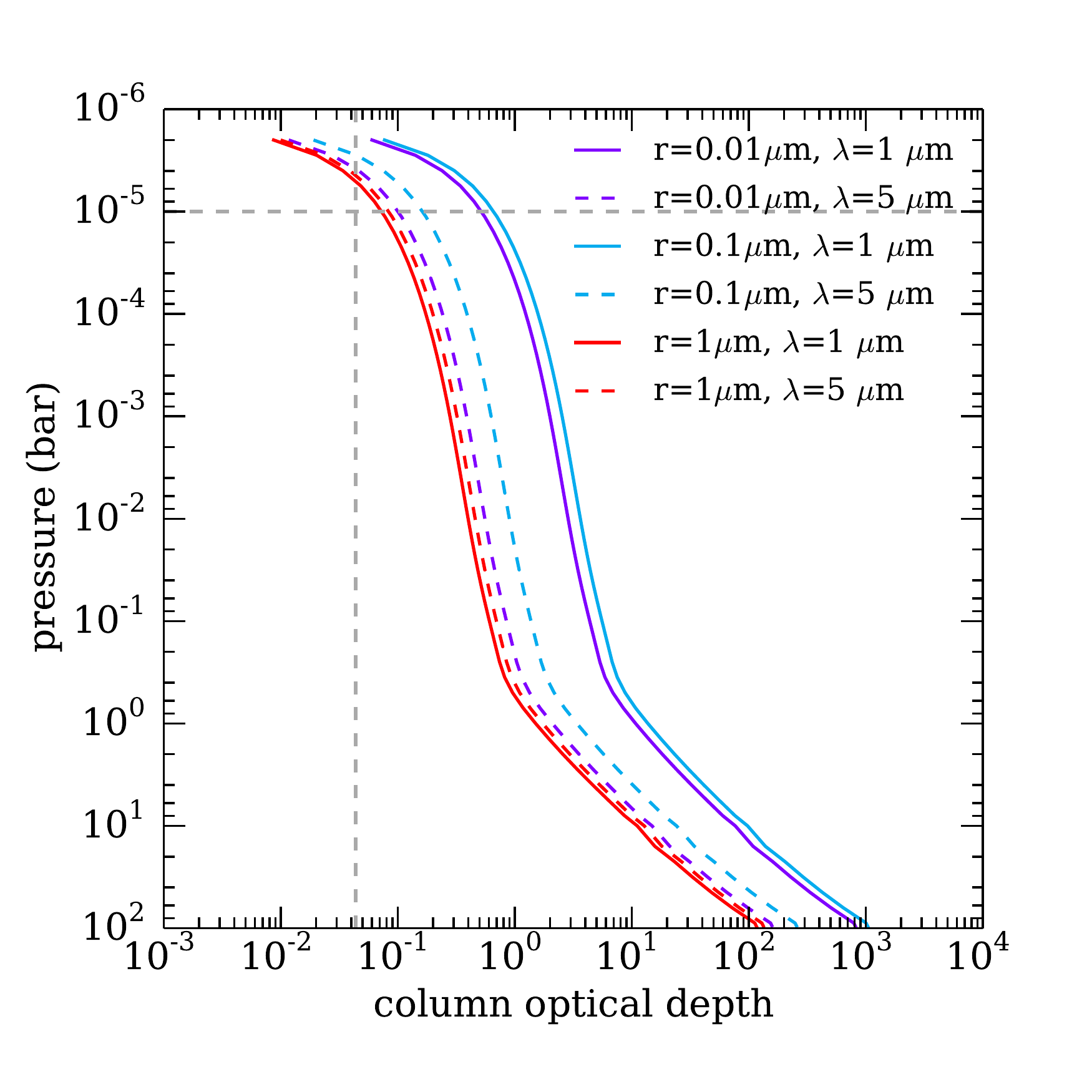}
 \caption{Column optical depth for hazes with varied radii (0.01 to 1\micron), 50$\times$ solar metallicity composition, \fhaze=10\%, and GJ 1214b's incident flux. Column optical depth is shown for two wavelengths (1 and 5 \micron) as a function of pressure. Note that smaller particles result in more wavelength-dependent optical depth. The dashed vertical gray line shows the $\tau=1$ line for slant viewing geometry using equation 6 from \ct{Fortney05c}. The dashed horizontal gray line shows the approximate altitude of GJ 1214b's cloud to cause a flat transmission spectrum.}
\label{photochem_taus}
\end{figure}

We calculate the abundances of soot precursors in the upper atmosphere using the photochemical model described extensively in \ct{Kempton12}, which is based on the methods published in \ct{Zahnle09a}. Briefly, the models use a chemical kinetics model to calculate disequilibrium chemistry due to both vertical mixing and photochemistry in the planetary atmosphere. The eddy diffusion coefficient, which parameterizes vertical mixing in the atmosphere, is taken as a free parameter that can be varied. We use the 50$\times$ solar metallicity results first published in \ct{Fortney13}, at five different irradiation levels (0.3, 1, 3, 10, 30$\times$ the true irradiation of GJ 1214b) and two eddy diffusion coefficients (\kzz= 10$^8$ and 10$^{10}$ cm$^2$ s$^{-1}$). We use the UV stellar spectrum measured by \ct{France13}. 

Figure \ref{photochem_single} shows the carbon chemistry in a single model as an example, at GJ 1214b's irradiation level and 50$\times$ solar composition. Because it is cool ($\sim$600 K), the atmosphere is dominated by methane at most altitudes. At the top of the atmosphere, methane is dissociated by UV flux from the host star. The chemistry that proceeds generates a variety of soot precursors (C$_2$H$_2$, C$_2$H$_4$, C$_2$H$_6$, C$_4$H$_2$, and HCN). These are the highest order hydrocarbons that can be generated with this particular model, as reactions to form higher-order hydrocarbon molecules in these environments are incompletely understood \cp[see, e.g., ][]{Moses11}. Nonetheless, unsaturated hydrocarbons like these soot precursors will continue to react and will likely form complex molecules (see, e.g. \ct{Yung84} and discussion of photochemical haze production in \ct{Morley13}). 

 \begin{figure}[t]
  \center   \includegraphics[width=3.4in]{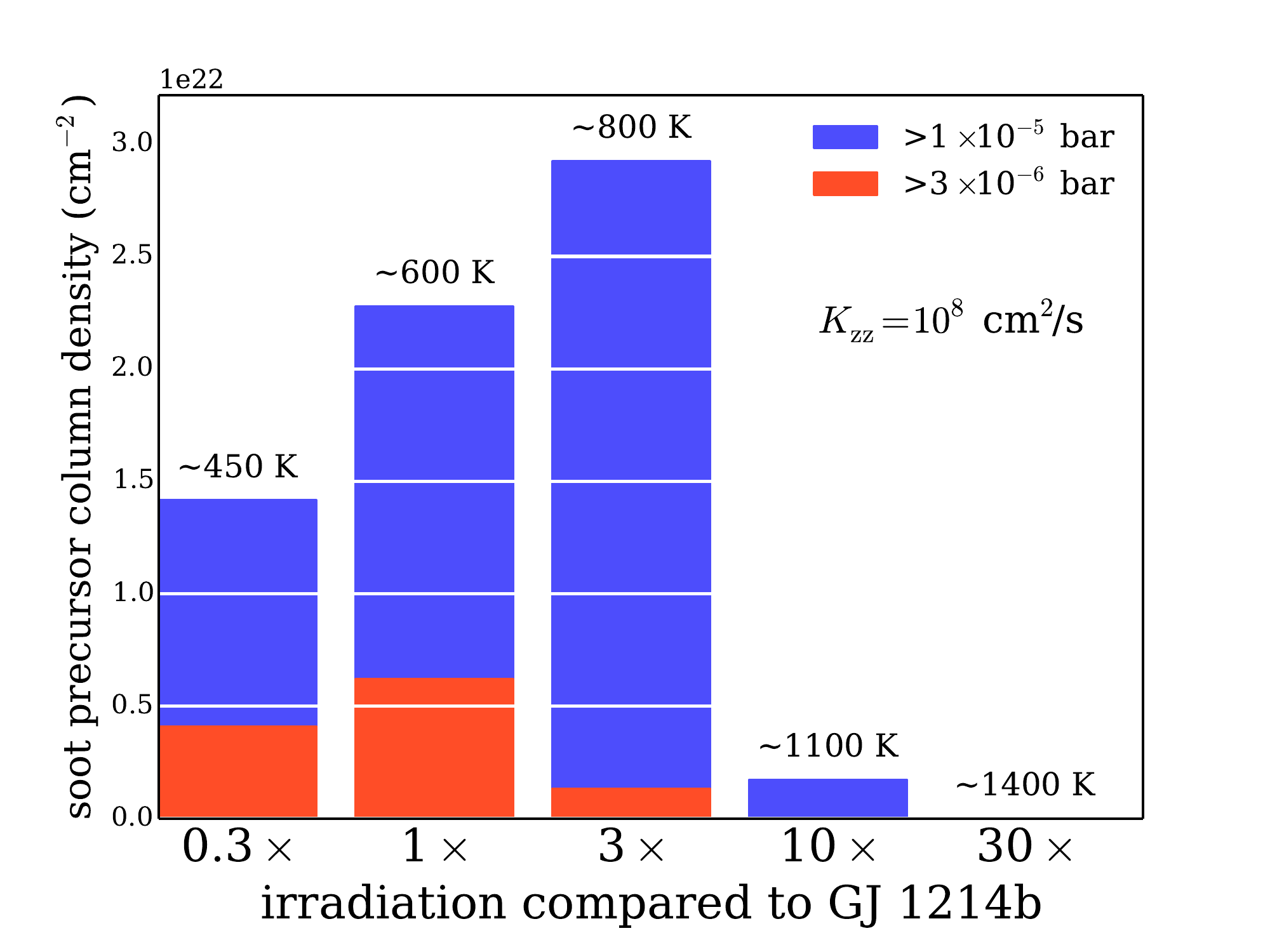}
	\vspace{-0.8cm}
  \center   \includegraphics[width=3.4in]{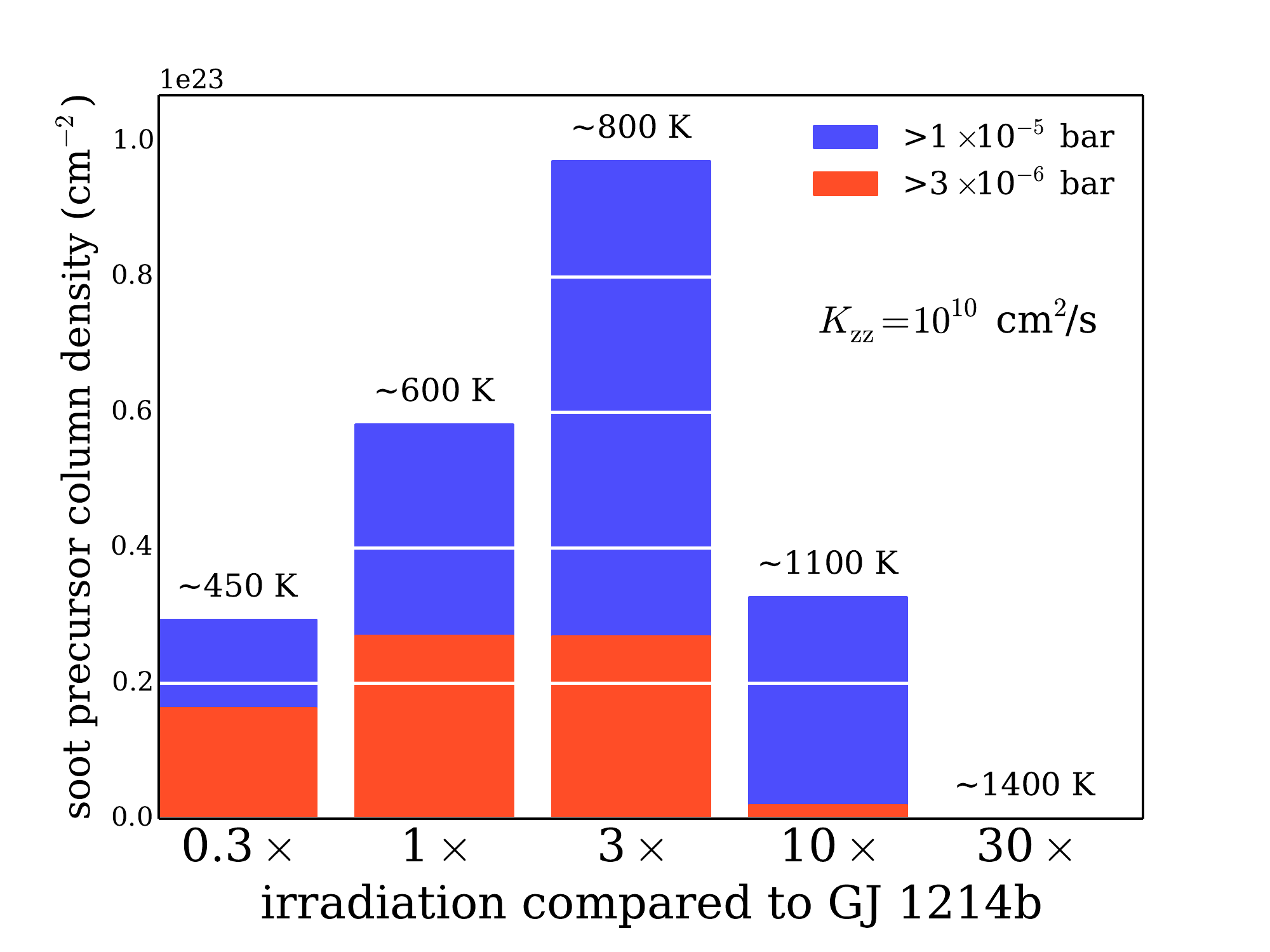}
 \caption{Summary of soot precursor production at high altitudes at 50$\times$ solar composition. The blue and red bars show the total mixing ratio of soot precursors above $10^{-5}$ and 3$\times10^{-6}$ bar respectively. Top panel shows \kzz=10$^8$ cm$^2$ s$^{-1}$; bottom panel shows \kzz=10$^{10}$ cm$^2$ s$^{-1}$. Models with high \kzz\ and 1--3$\times$ the irradiation of GJ 1214b have the most soot precursors. }
\label{photochem_bar}
\end{figure}

Figure \ref{photochem_summed} illustrates how both \kzz\ and incident flux affect the formation of these soot precursors. The mixing ratios of C$_2$H$_2$, C$_2$H$_4$, C$_2$H$_6$, C$_4$H$_2$, and HCN are summed at each layer of the model. As found in \ct{Fortney13} using the same models, we find that models with 1--3$\times$ GJ 1214b's irradiation have the most soot precursors at high altitudes. In the hotter, high irradiation models (20--30$\times$), the atmosphere is dominated by CO instead of CH$_4$; the chain of chemistry that begins with methane dissociation cannot start in a CO dominated atmosphere, as CO's bond is less easily broken with UV light. The lower production of soot precursors at low irradiation levels is because the rate of methane dissociation is lower. The  production of soot precursors can also be a strong function of the eddy diffusion parameter \kzz; this is especially true at temperatures that are close to the boundary between CO and CH$_4$ dominated atmospheres (20$\times$, $\sim$1200 K), because the vigor of vertical mixing changes the bulk carbon chemistry. 

Figure \ref{photochem_taus} shows the haze column optical depth for three example models, each with 50$\times$ solar metallicity, \fhaze=10\%, and GJ 1214b's incident flux. Three different particle sizes spanning our model grid are shown, and the column optical depth is calculated for two wavelengths spanning the infrared (1 and 5 \micron). We find that 1 \micron\ particles have the lowest optical depth and relatively constant optical depth across the infrared. 0.1 and 0.01 \micron\ particles have more wavelength dependent optical depth, as expected for small particles.

Figure \ref{photochem_bar} summarizes these findings. We calculate the column density of the soot precursors in high altitude layers of the model (above $10^{-5}$ and $3\times10^{-6}$ bar). We find that the largest quantity of soot precursors are in models with high \kzz\ and 1--3$\times$ GJ 1214b's irradiation level. 

GJ 1214's stellar spectrum is used for all photochemical calculations, so we note that the results will depend on the UV spectrum of the host star, even with the same total incident flux.

 \begin{figure*}[tb]
     \includegraphics[width=7.3in]{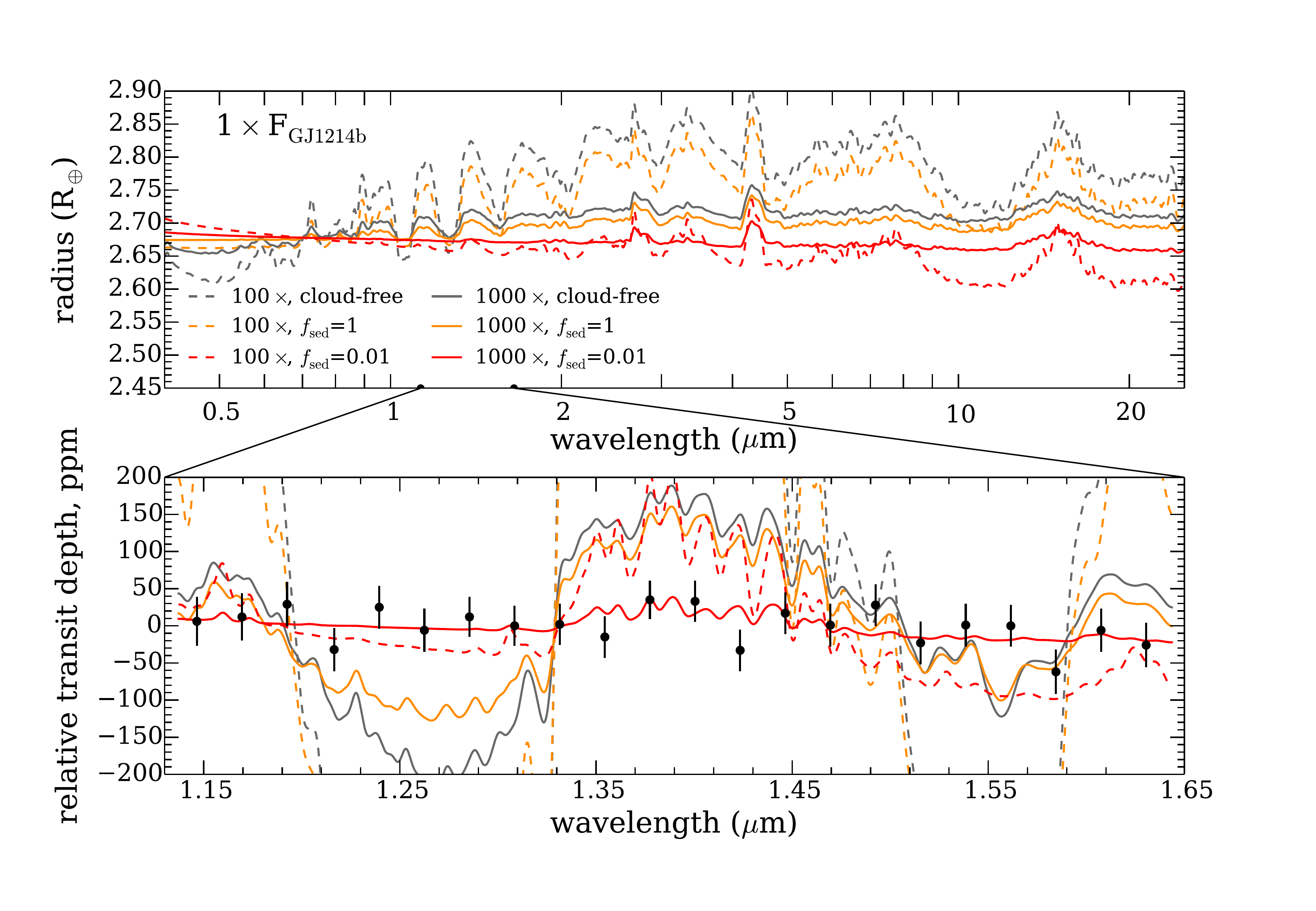}
 \caption{Example high metallicity (100 and 1000$\times$ solar) transmission spectra with and without clouds. The top panel shows the optical and infrared transmission spectra. The bottom panel shows the same spectra,  zoomed in to focus on the \ct{Kreidberg14} data in the near-infrared. Cloud-free transmission spectra are shown as light and dark gray lines and cloudy spectra are shown as colored lines. Note that the only model that fits the data is the 1000$\times$ solar model with \fsed=0.01 (lofted) clouds. }
\label{trans_specs_fsed001}
\end{figure*}

 \begin{figure*}[h]
     \includegraphics[width=3.75in]{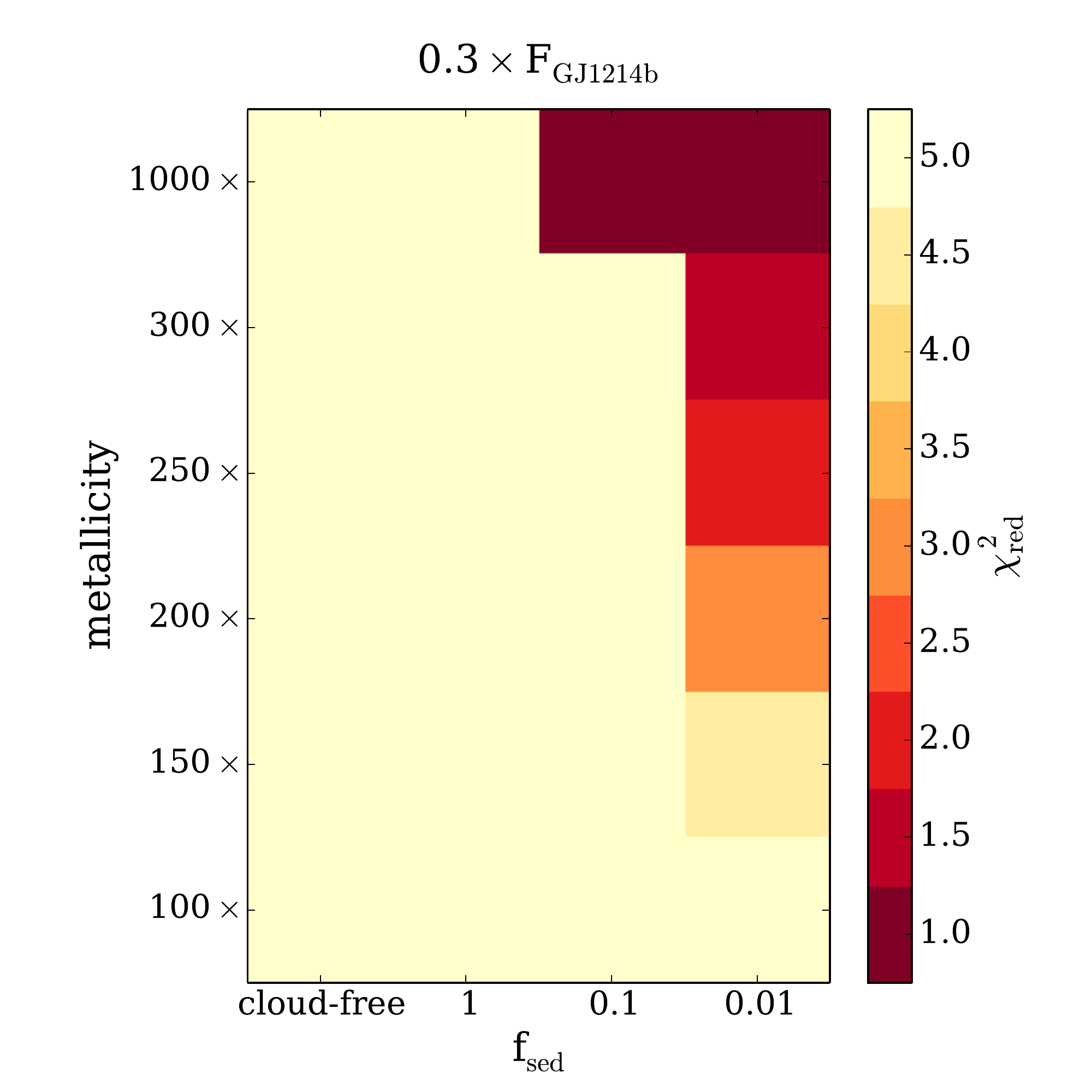}
     \includegraphics[width=3.75in]{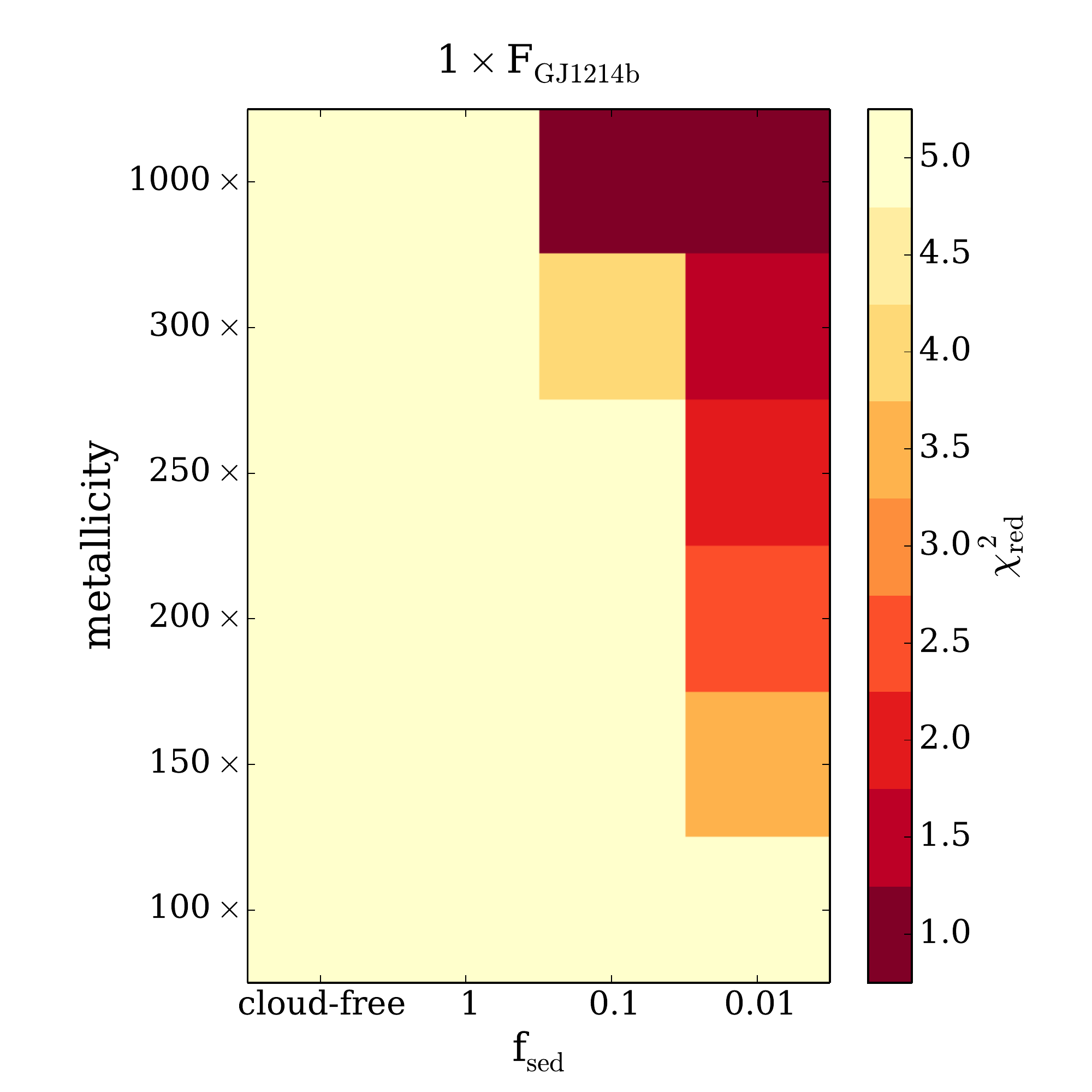}
     \includegraphics[width=3.75in]{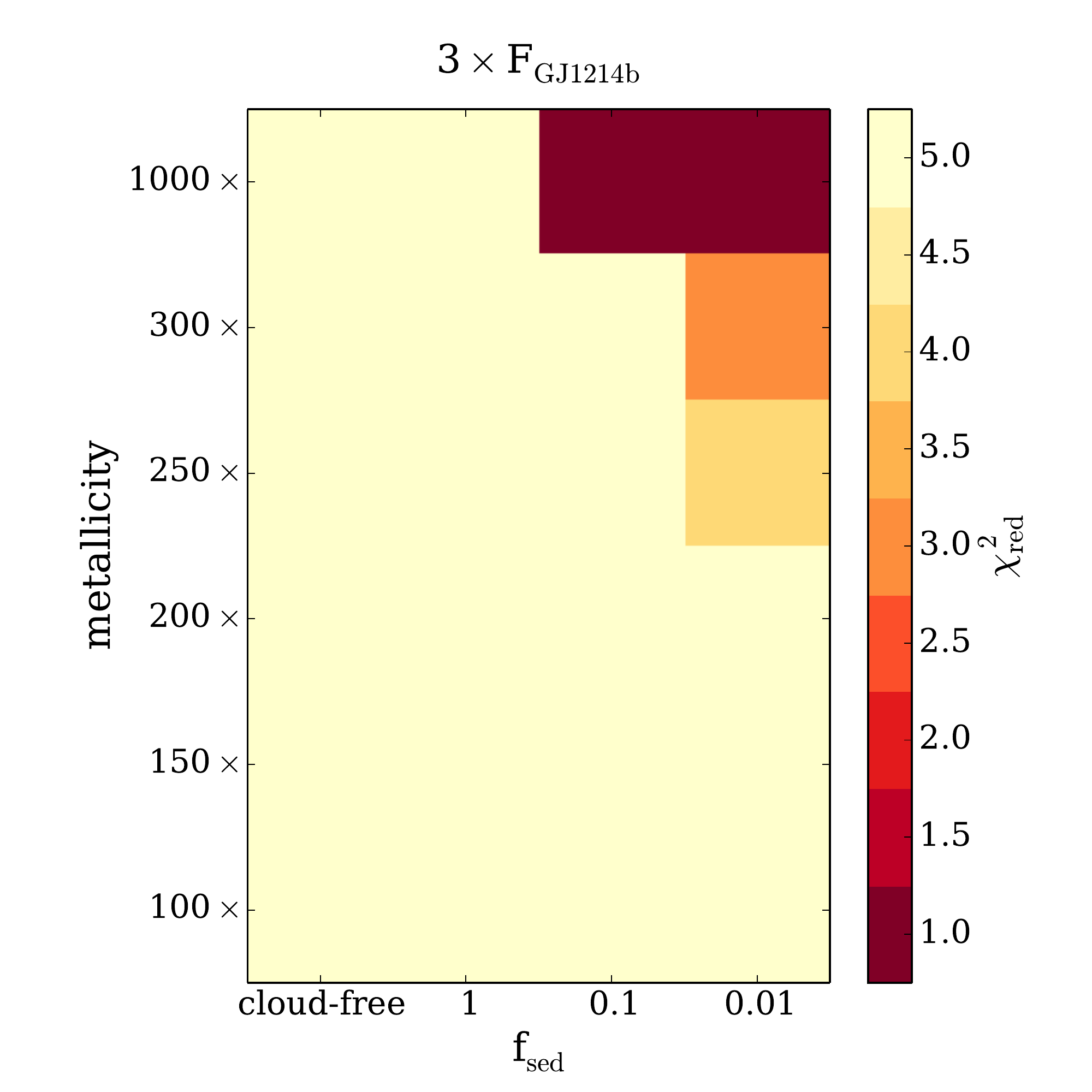}
     \includegraphics[width=3.75in]{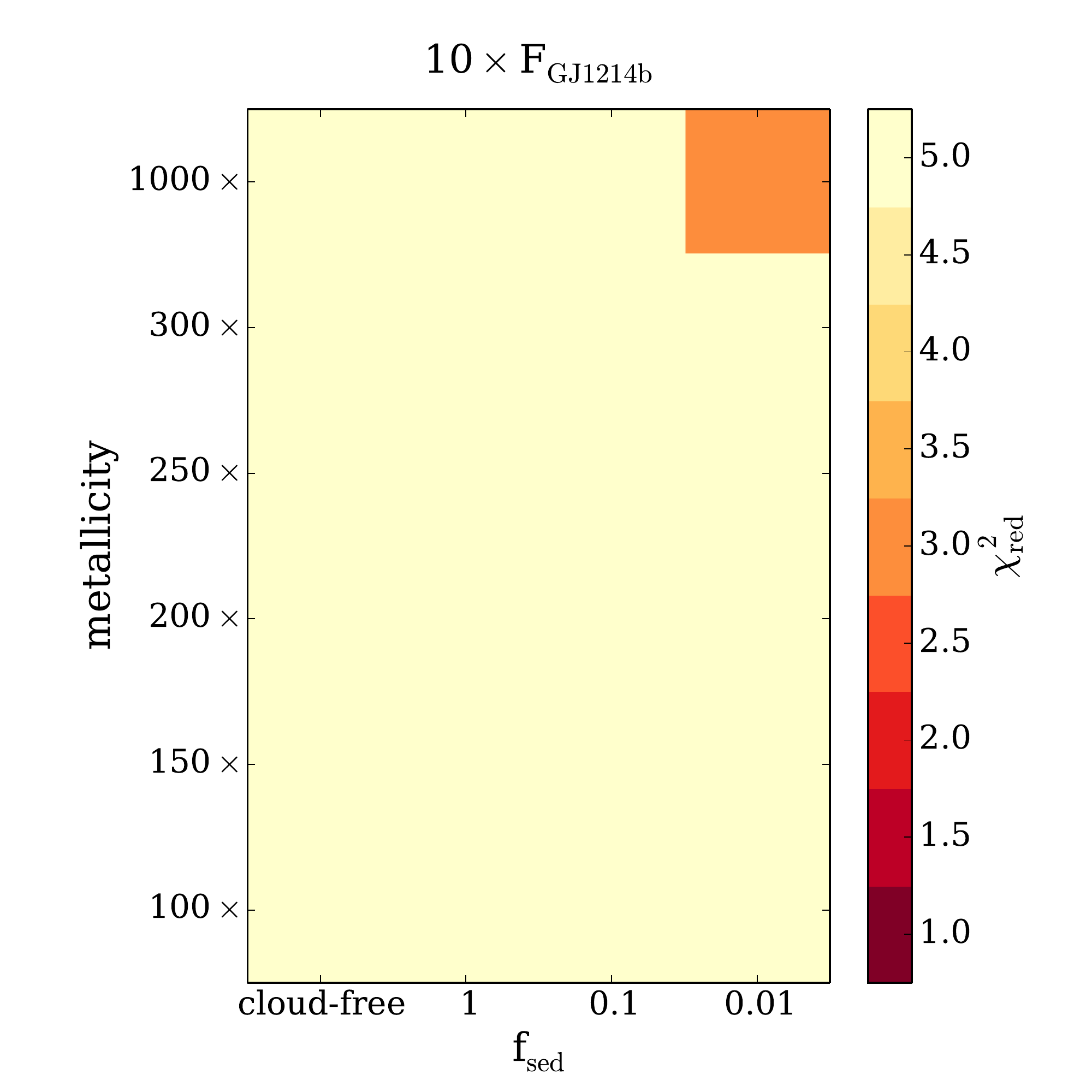}
 \caption{Chi-squared maps showing quality of fit to \ct{Kreidberg14} data for transmission spectra with equilibrium clouds, with varied irradiation levels, metallicities, and cloud sedimentation efficiency \fsed. Starting in the top left panel, models with 0.3, 1, 3, and 10 $\times$ GJ 1214b's irradiation are shown. Dark red sections show acceptable fits (reduced $\chi^2$ close to 1.0). Note that high metallicity and low \fsed\ (lofted clouds) are simultaneous requirements for these clouds to generate a flat enough transmission spectrum to be consistent with the data. }
\label{chisquare_map_eq}
\end{figure*}

\subsubsection{Photochemical hazes} \label{makinghazes}

We follow the approach developed in \ct{Morley13} to calculate the locations of soot particles based on the results from the photochemical models. We sum the densities of the five soot precursors (C$_2$H$_2$, C$_2$H$_4$, C$_2$H$_6$, C$_4$H$_2$, and HCN) to find the total mass in soot precursors. We assume that the soots form at the same altitudes as the soot precursors exist: we multiply the precursors' masses by our parameter $f_{haze}$ (the mass fraction of precursors that form soots) to find the total mass of the haze particles in a given layer. For each layer,

\begin{equation}
M_{haze} = f_{haze} \times(M_{C_2H_2} +M_{C_2H_4} +M_{C_2H_6} +M_{HCN} + M_{C_4H_2}) 
\end{equation}
where $f_{haze}$ is the efficiency, $M_x$ is the mass of material in each species within each model layer from the photochemical model, and $M_{haze}$ is the calculated mass of haze particles in each layer.

We vary both $f_{haze}$ and the mode particle size (assuming a log-normal particle distribution); we calculate the number of particles by summing over the distribution for each of our chosen particle sizes. Soot optical properties (the real and imaginary parts of the refractive index) from the software package OPAC (Optical Properties of Aerosols and Clouds) (Hess et al. 1998), were used and linearly extrapolated in wavelength for wavelengths longer than 40 \micron.

\subsection{Transmission Spectra}

We calculate the transmission spectrum for each converged P--T profile, including the effect of clouds. The optical depths for light along the slant path through the planet's atmosphere are calculated at each wavelength, generating an equivalent planet radius at each wavelength. The model is extensively described in \ct{Fortney03} and \ct{Shabram11}. Cloud layer cross-sections generated from the model atmosphere are treated as pure absorption, and are added to the wavelength-dependent cross-sections of the gas.

 \begin{figure*}[tb]
 \vspace{-1cm}
  \hspace{-1.5cm}
\begin{minipage}[t] {\textwidth}
    \includegraphics[width=8.5in]{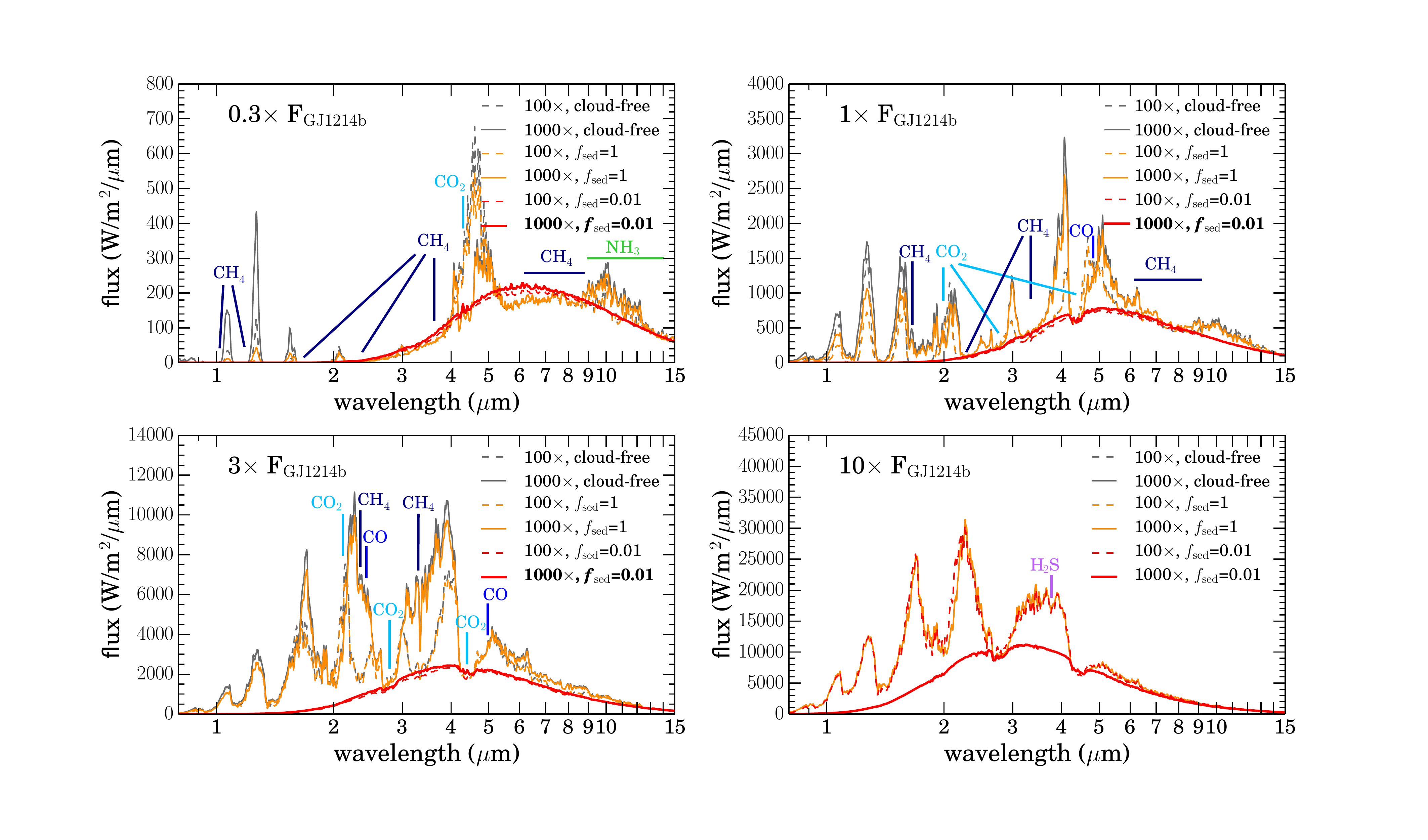}
  \end{minipage} 
  \vspace{-0.5cm}
 \caption{Thermal emission spectra of models with sulfide and salt clouds. Each panel shows models with a different incident flux. Gray lines show cloud-free models and colored lines show cloudy models. The fonts in the captions are bolded if the transmission spectrum with those parameters fits the \ct{Kreidberg14} data. For the cooler models, the cloud opacity decreases the near-infrared flux. For the warmer models, the clouds are optically thinner. Major molecular features are labeled. Unlabeled major features are predominantly H$_2$O. }
\label{thermal_eq}
\end{figure*}

 \begin{figure*}[tb]
  \vspace{-.5cm}
  \hspace{-1.5cm}
\begin{minipage}[t] {\textwidth}
     \includegraphics[width=8.5in]{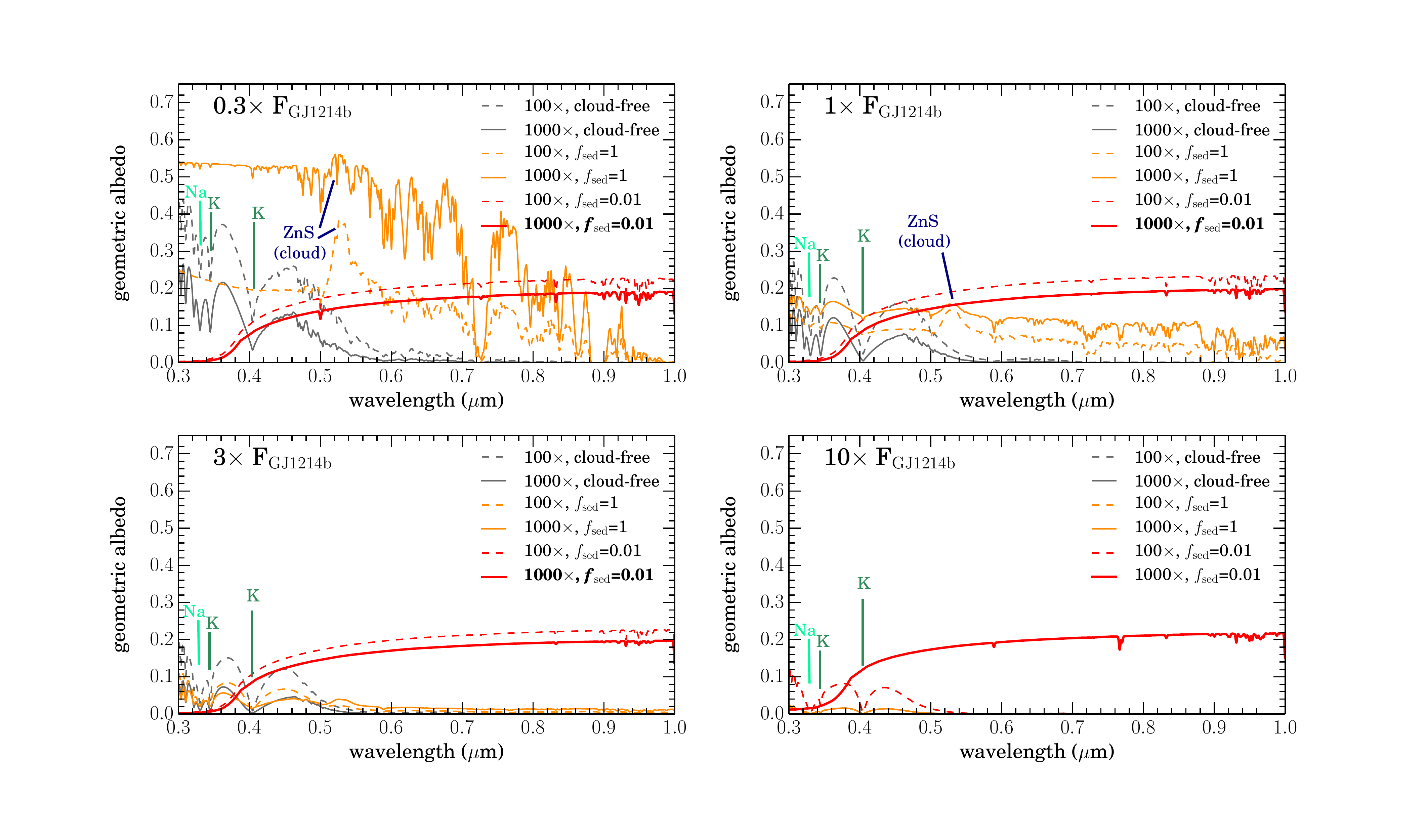}
 \end{minipage} 
 \vspace{-.5cm}
 \caption{Albedo spectra for models with salt/sulfide clouds. The top set of panels show thinner clouds (\fsed=1) and the bottom set of panels show thicker clouds (\fsed=0.01). Bolded legend text indicates models that fit the transmission spectrum data. Each panel shows a different incident flux compared to GJ 1214b. }
\label{alb_specs_eq}
\end{figure*}

\subsection{Thermal Emission Spectra }

A new model to calculate the thermal emission of a planet with arbitrary composition and clouds was developed for this work. The model includes absorption and scattering from molecules, atoms, and clouds. We use the C version of the open-source radiative transfer code \texttt{disort} \cp{Stamnes88, Buras11} which uses the discrete-ordinate method to calculate intensities and fluxes in multiple-scattering and emitting layered media. We describe this new calculation in more detail in the Appendix.

\subsection{Albedo Spectra}

We calculate reflected light spectra of each model atmosphere using the methods developed for planets and described in detail in \ct{Toon77, Toon89, Mckay89, Marley99, Marley99b, Cahoy10}. Here, we use the term geometric albedo to refer to the albedo spectrum at full phase ($\alpha$=0, where the phase angle $\alpha$ is the angle between the incident ray from the star to the planet and the line of sight to the observer): 

\begin{equation}
A_g(\lambda)=\dfrac{F_p(\lambda,\alpha=0)}{F_{\odot,L}(\lambda)}
\end{equation}    
where $\lambda$ is the wavelength, $F_p(\lambda,\alpha=0)$ is the reflected flux at full phase, and $F_{\odot,L}(\lambda)$ is the flux from a perfect Lambert disk of the same radius under the same incident flux. 

The absorption and scattering properties of clouds are calculating using Mie theory, assuming homogeneous, spherical particles.

\section{Results: Equilibrium Clouds}  \label{eq_clouds}

A grid of 96 models with salt and sulfide clouds (ZnS, KCl, \nas) are calculated, with irradiations of 0.3, 1, 3, and 10$\times$ GJ 1214b's, metallicities of 100, 150, 200, 250, 300, and 1000$\times$ solar, and \fsed\ of 1, 0.1, 0.01, and cloud-free. A smaller grid of cold models with water clouds are calculated, with 0.01, 0.03, and 0.1$\times$ GJ 1214b's incident flux, 50, 300, and 1000$\times$ solar metallicity, and \fsed\ of 1, 0.1, 0.01, and cloud-free. For each of these sets of parameters, we calculate the transmission spectrum, thermal emission, and albedo spectrum; a representative sample of these models are shown in this section as well as summaries of their properties. The spectra are all available online at the lead author, Caroline Morley's, website, currently at http://www.ucolick.org/$\sim$cmorley.

\subsection{Transmission Spectra}

The top panel of Figure \ref{trans_specs_fsed001} shows examples of models at 1$\times$ GJ 1214b's irradiation and with metallicities of 100 and 1000 $\times$ solar, both with and without cloud opacity. The full grid also includes models at different temperatures (irradiation) and with intermediate and lower metallicities. 

For cloud-free models, transmission spectra have visible features from various atoms and molecules; the prominence of those features changes with both temperature (irradiation) and metallicity. For example, the alkali metals (Na, K) create the strongest features in the warmest (10$\times$ GJ 1214b's irradiation) models. As they condense into clouds in cooler planets, they become significantly less prominent. Other visible features include the major absorbers H$_2$O, CH$_4$, and CO. The size of features decreases at higher metallicities because the mean molecular weight increases, decreasing the scale height. As discussed in the introduction, the size of features is proportional to the scale height. The temperature of the atmosphere also controls the carbon chemistry; CO and CO$_2$ features dominate the mid-infrared spectrum at $10\times$ GJ 1214b's irradiation, whereas CH$_4$ dominates at 0.3$\times$.

We find that all clouds flatten the transmission spectrum, reducing the size of the features caused by molecules and atoms. The lowest \fsed\ values (indicating lofted clouds of small particles, as shown in Figure \ref{cloudcode}) flatten the spectrum the most because they become optically thick above the gas absorbers. Higher metallicity models have flatter spectra both because they have smaller scale heights (as seen in the cloud-free spectra as well) and a larger abundance of metals to form clouds, leading to optically thicker clouds.

\subsubsection{Comparing to the Kreidberg et al. 2014 data}

We compare all of the synthetic transmission spectra to the observations published in \ct{Kreidberg14}. These data are the highest signal-to-noise (SNR) spectra that have been obtained for this planet. A chi-squared analysis allows us to assess the relative goodness-of-fit for each model. We compare the hotter and colder models to the same observed data; since the data is consistent with a flat line, it represents our fiducial high SNR ``flat'' spectrum to explore the range of parameter space that is likely to have planets with featureless spectra. We note that we are not suggesting that GJ 1214b has a different incident flux than reality; we are using the observed data as a generic dataset representing a featureless spectrum. 

Examples of these fits are shown in the lower panel of Figures \ref{trans_specs_fsed001}. It is clear both by eye and using a chi-squared analysis that neither of the \fsed=1 models (thinner clouds) fit the data; the features in the models are significantly larger than the error bars or scatter in the data points. For the thicker clouds (\fsed=0.01) only the highest metallicity model matches the data well. 

These results are summarized across the entire modeled parameter space in Figure \ref{chisquare_map_eq}. We calculate reduced $\chi^2$ assuming 20 degrees of freedom (22 data points $-$ 2 fitted parameters). We consider acceptable fits to be those with $\chi^2_{\rm red}<1.14$, corresponding to P=0.3 of exceeding $\chi^2$ assuming 20 degrees of freedom \cp{Bevington03}. In Figure \ref{chisquare_map_eq}, the dark red regions represent the lowest reduced $\chi^2$. We find that only models at low \fsed\ and very high metallicity ($\sim$1000 $\times$ solar) can flatten the transmission spectrum enough to match the data. We assess whether this corner of parameter space is likely in Sections \ref{highmetlikely} and \ref{lowfsedlikely}.

\subsection{Thermal Emission Spectra}

Figure \ref{thermal_eq} shows thermal emission spectra for models with thin and thick clouds. The cloud-free models are dominated by features from water, methane, and carbon monoxide. As in the transmission spectra, warmer objects have deeper CO features and cooler objects have deeper CH$_4$ features. Note that at 3$\times$ GJ 1214b's irradiation ($\sim$800 K) the amount of methane is strongly metallicity dependent. Lower metallicity models (100$\times$ solar) show a deep methane features between 2 and 4 \micron, whereas higher metallicity models have a shallower feature. 

Thin (\fsed=1) clouds marginally change the thermal emission. The difference is very small at 3--10$\times$ GJ 1214b's irradiation. For the cooler two sets, the clouds decrease the flux in the near-infrared (0.8--2 \micron) but leave longer wavelengths unchanged. 

Thick clouds (\fsed=0.01)---the value of \fsed\ needed to flatten the spectrum to match observations---dramatically change the thermal emission. At all temperatures, the planet has fewer features and a smoother spectrum. This difference is because the clouds create an optically thick layer, blocking the passage of photons from deeper, hotter layers in the atmosphere. Essentially, we are seeing an optically thick, relatively gray, cloud layer.

\subsection{Albedo Spectra}

Albedo spectra at each irradiation level are shown in Figure \ref{alb_specs_eq}. As in hot Jupiter models \cp[e.g.][]{Sudarsky00}, at these high metallicities and warm temperatures, the albedo spectra of these objects will be very dark, especially at wavelengths beyond 0.6 \micron. In particular, the alkali metals create strong absorption features that carve away the reflected light. 

For models with thin clouds (\fsed=1) at 0.3--1$\times$ GJ 1214b's irradiation, the clouds brighten the albedo spectra at most wavelengths. Absorption features from methane, alkalis, and water are visible. A feature from the reflection of spherical ZnS particles is clearly visible in the models at 0.53 \micron. This reflection feature depends on the particle size distribution in the cloud: larger particles ($>$3--5 \micron) create a larger feature. Warmer models (3--10$\times$ GJ 1214b's irradiation) with thin clouds lack these interesting cloud features and have lower albedos; the clouds are too deep in the atmosphere to change the albedo spectra significantly.  

Reflection spectra of models with thick clouds (\fsed=0.01) look significantly different. The scattering properties and locations of the clouds substantially change reflected light from a planet. For these models with thick clouds, they are made of small particles highly lofted in the atmosphere (see Figure \ref{cloudcode}). They absorb efficiently at wavelengths from 0.3--0.5 \micron\ and scatter more efficiently beyond 0.5 \micron, creating a spectrum that slopes upward to red wavelengths. Some absorption from water vapor between 0.9 and 1.0 \micron\ are visible, but most of the gas absorption features seen in less cloudy models are muted.

\subsection{Cold Planets with Water Clouds}

Measuring reflected light using optical secondary eclipse depths will be extremely challenging for small, cool planets like GJ 1214b. A set of small planets that may actually be more accessible for reflected light spectroscopy will be directly-imaged distant companions, observed with telescopes like the Wide-Field Infrared Survey Telescope (\emph{WFIRST}) or another dedicated space-based coronagraphic telescope \cp{Spergel15}. These planets will be colder, more like the planets in our own solar system ($\sim$50--300 K). 

These planets may be more accessible in part because many of them will have condensed volatile clouds in their atmospheres, like water, ammonia, and methane. These volatile clouds have higher single scattering albedos in the optical compared to refractory clouds like salts, sulfides, and silicates. The importance of clouds in increasing the albedo at red and far red wavelengths was noted by \ct{Marley99} and \ct{Sudarsky00}. 

\begin{figure}[t]
 \hspace{-5mm}
     \begin{minipage}[t] {0.5\textwidth}
	     \includegraphics[width=3.8in]{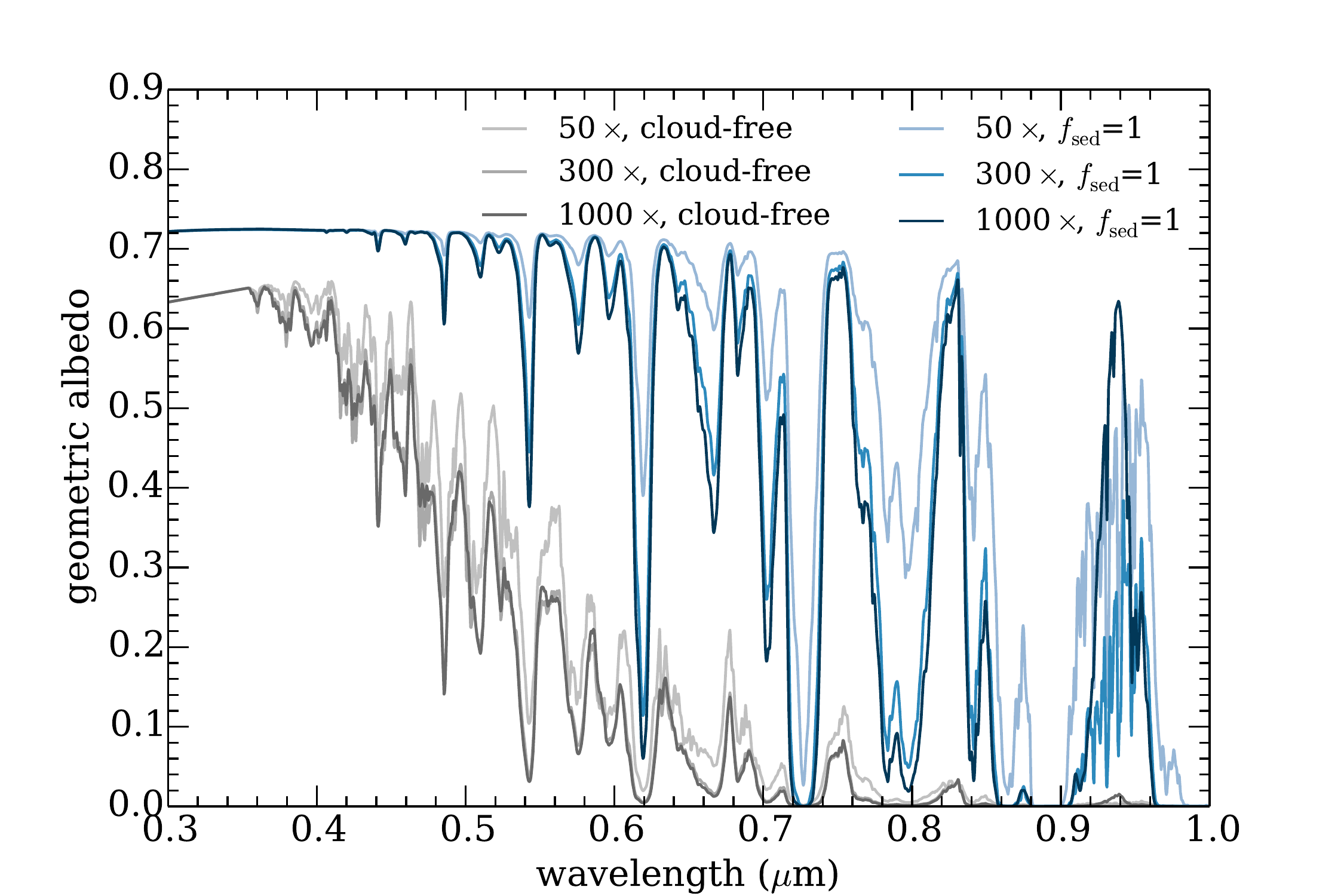}
	     \includegraphics[width=3.8in]{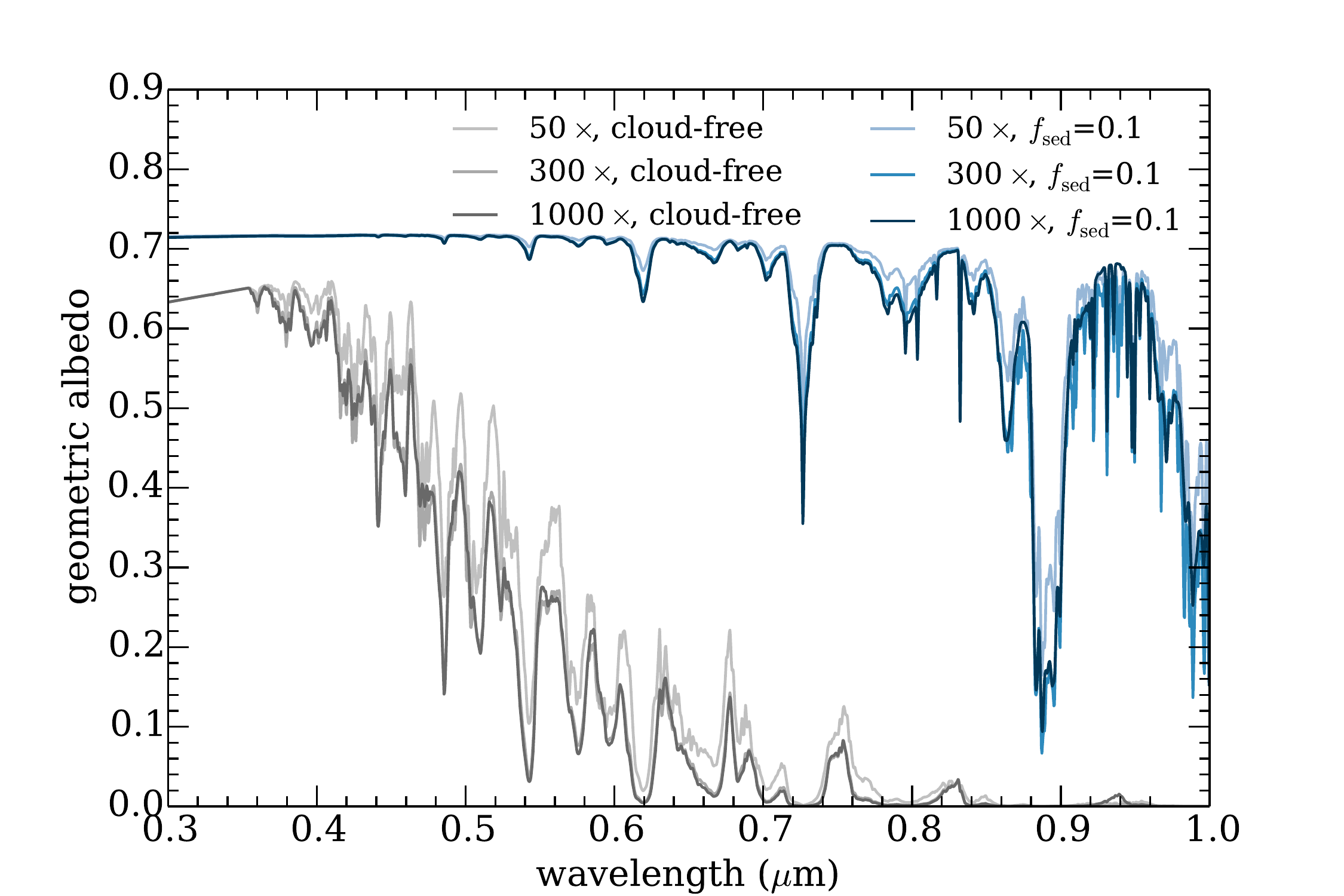}
	     \includegraphics[width=3.8in]{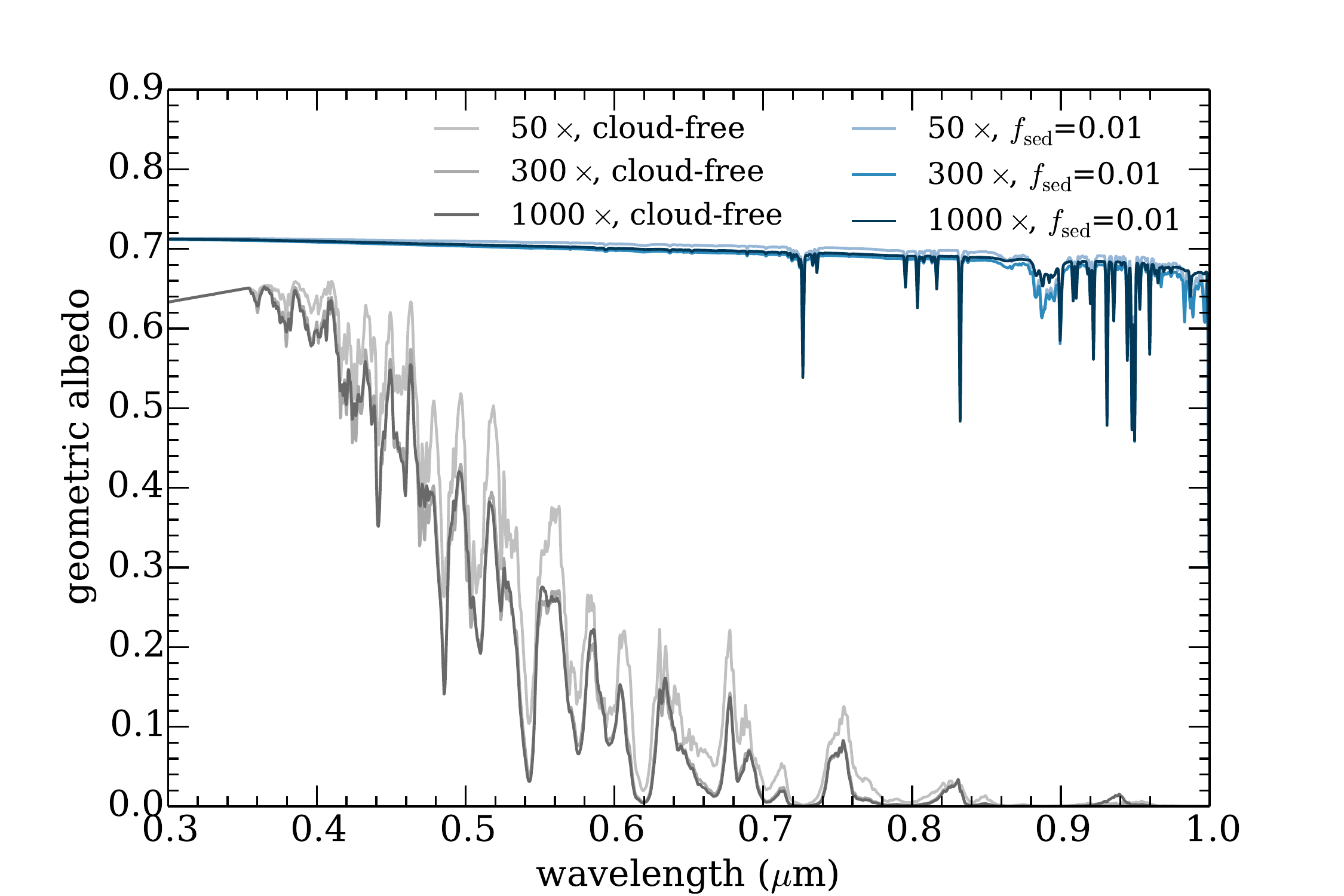}
      \end{minipage}
 \caption{Albedo spectra for cold models (\teff=190 K) with water clouds at 50--1000$\times$ solar metallicity. The top, middle, and bottom panels show models with \fsed=1, 0.1, and 0.01 respectively. Note that water clouds create bright albedo spectra with strong features from methane.  }
\label{alb_specs_cold}
\end{figure}

Cold ($\sim$200 K) reflected light spectra for small planets with enhanced metallicity atmospheres are shown in Figure \ref{alb_specs_cold}. In the absence of clouds, planets are predicted to be bright at short wavelengths ($\sim$0.3--0.6 \micron) due to efficient Rayleigh scattering at short wavelengths and fainter from 0.6 to 1 \micron. The features are mostly caused by methane absorption.

Spectra with ice clouds are significantly brighter at all wavelengths. The \fsed=1 models (thinner clouds) have large features caused mostly by methane absorption bands of varying strengths. Some water absorption features are also visible from 0.9--1 \micron. In our parameterization, an \fsed\ value of 1--3 is consistent with Jupiter's ammonia clouds \cp{AM01}, so it is reasonable to imagine that cold, old exoplanets will have similar clouds. 

For thicker clouds (\fsed=0.1 and 0.01) the planet becomes more uniformly bright; this change is because the clouds reflect light at higher altitudes than photons are absorbed by molecules, except within the strongest methane bands (e.g. at 0.88 \micron). Bright high altitude clouds would make planets detectable, but challenging to characterize since they have fewer molecular features.

\section{Results: Photochemical Hazes} \label{phot_hazes}

We consider a grid of 100 models with irradiation of 0.3, 1, 3, 10, and 30$\times$ GJ 1214b's, $f_{haze}$ of 1, 3, 10, and 30\%, and mode particle sizes of 0.01, 0.03, 0.1, 0.3, and 1 \micron, and optical properties of soot, as described in Section \ref{makinghazes}. All models have compositions of 50$\times$ solar metallicity. 
 
\subsection{Temperature Structure and Anti-greenhouse Effect} \label{antigreen}

Unlike the equilibrium cloud models, for the models with photochemical hazes, the temperature structure is calculated self-consistently with the haze opacity (though the photochemistry to calculate the abundance of soot precursors is calculated using a constant haze-free temperature profile, see Section \ref{methods}).  

For models that contain dark soot particles at high altitudes, these particles are efficient optical absorbers and heat the upper layers of the atmosphere. This phenomenon is called the ``anti-greenhouse effect" and has been well-documented in solar system atmospheres. For example, Titan's atmosphere is exactly analogous: a photochemical haze at high altitudes creates a temperature inversion and cools Titan's surface \cp{McKay91, McKay99}. 

\begin{figure}[tb]
     \includegraphics[width=3.7in]{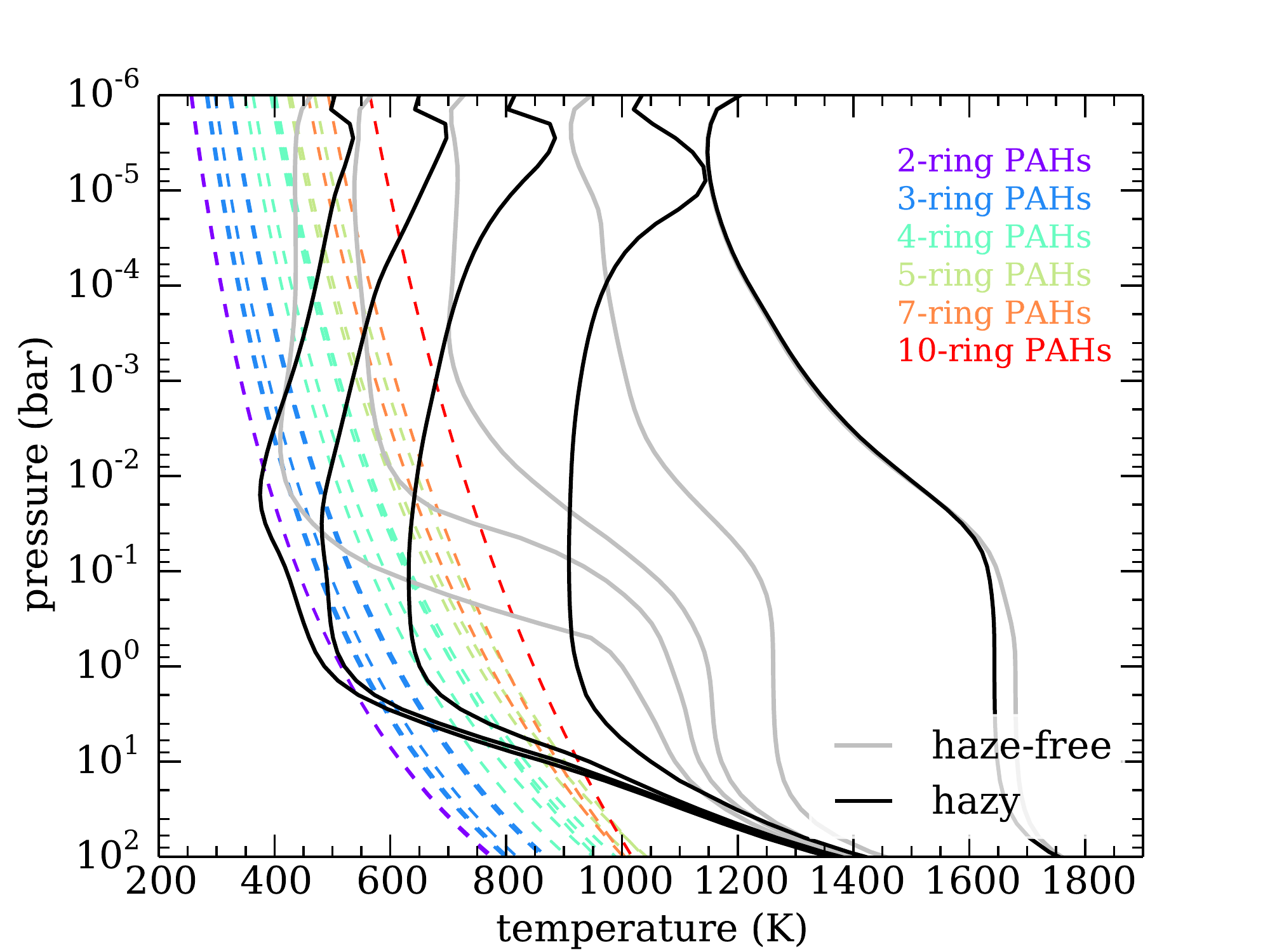}
 \caption{Pressure--temperature profiles of clear and hazy models are shown as gray and black lines, respectively. From left to right, these models have irradiation levels of 0.3, 1, 3, 10, and 30 times GJ 1214b's. The hazy models have particle sizes of 0.1 \micron\ and \fhaze=10\%. The colored dashed lines show the condensation temperatures of a number of different polycyclic aromatic hydrocarbons (PAHs), color-coded by the size of the molecule.}
\label{pah_cond}
\end{figure}

Figure \ref{pah_cond} shows this effect for our grid of hazy models. The gray lines show haze-free temperature profiles of GJ 1214b analogs from 0.3 to 30 $\times$ GJ 1214b's irradiation. The black lines show models with hazes in their upper atmospheres. The haze particles absorb more efficiently at optical wavelengths than they do in the infrared; this means that they absorb stellar flux but allow the thermal flux from deeper layers to escape. The absorption from hazes means that less stellar flux reaches deeper parts of the atmosphere. Since the upper atmosphere has a low infrared emissivity, in order to radiate the energy from the absorbed stellar flux, the upper layers must reach higher temperatures.

\subsection{Molecular Size of Condensible Hydrocarbons}

The temperatures at which various hydrocarbons evaporate are also shown in Figure \ref{pah_cond}. These boiling temperatures ($T_{\rm evap}$) are calculated using the lab-measured values of the boiling point at standard temperature and pressure ($T_{\rm STP}$) and the enthalpy of vaporization ($\Delta H_{\rm vap}$). These are related by the Clausius-Clapeyron relationship for a phase change at constant temperature and pressure, 
\begin{equation}
T_{\rm evap} = \left[\dfrac{1}{T_{\rm STP}}-\dfrac{R \ln \frac{P}{P_{\rm STP}} } {\Delta H_{\rm vap}}\right] ^{-1}. 
\end{equation}

These curves look similar to condensation curves (as shown in Figure \ref{ptprofs_eq}), but are physically not the same. The boiling temperature here represents the boundary where it is possible to have solid or liquid material in the atmosphere. This value is not the same as the condensation curve, which represents the point in temperature and pressure at which an atmosphere with a certain composition (usually assuming equilibrium chemistry) has a vapor pressure of that material equal to the saturation vapor pressure. 

Boiling temperatures are calculated for polycyclic aromatic hydrocarbons (PAHs) that range in size from two aromatic rings to ten. Specifically we include Azulene, 1-Methylnaphthalene (2 rings), Anthracene, Acenaphthene, Acenaphthylene, Phenanthrene, Fluorene (3 rings), Chrysene, Benz[a]anthracene, Fluoranthene, Pyrene, Triphenylene (4 rings), Dibenz[a,h]anthracene, Benzo[k]fluoranthene, Benzo[a]pyrene (5 rings), Benzo[ghi]perylene, Coronene (7 rings), and Ovalene (10 rings). The laboratory data for these PAHs were found using the NIST database (http://webbook.nist.gov/). 

We find that, as expected, larger hydrocarbons boil at higher temperatures than smaller hydrocarbons. As noted in \ct{Liang04}, small hydrocarbons (including many of the PAHs shown here) will not be able to condense in warm planetary atmospheres. 

This conclusion has a few implications. To have condensed haze material in a $\sim$600 K atmosphere like GJ 1214b's, the 2--4 carbon soot precursors (produced in the photochemistry model) must react many more times to make 10 or more ring PAHs, or other equivalently large hydrocarbons. We can expect that some of these intermediate materials---which must be vapor at these temperatures---are likely to be present in these atmospheres. If we could characterize the composition of vapor PAHs---the building blocks of hazes---in a hazy atmosphere, we could constrain the chemical pathways to form the condensed hazes. 

 \begin{figure*}[tb]
 	\hspace{-0.5cm}
   \center  \includegraphics[width=7.3in]{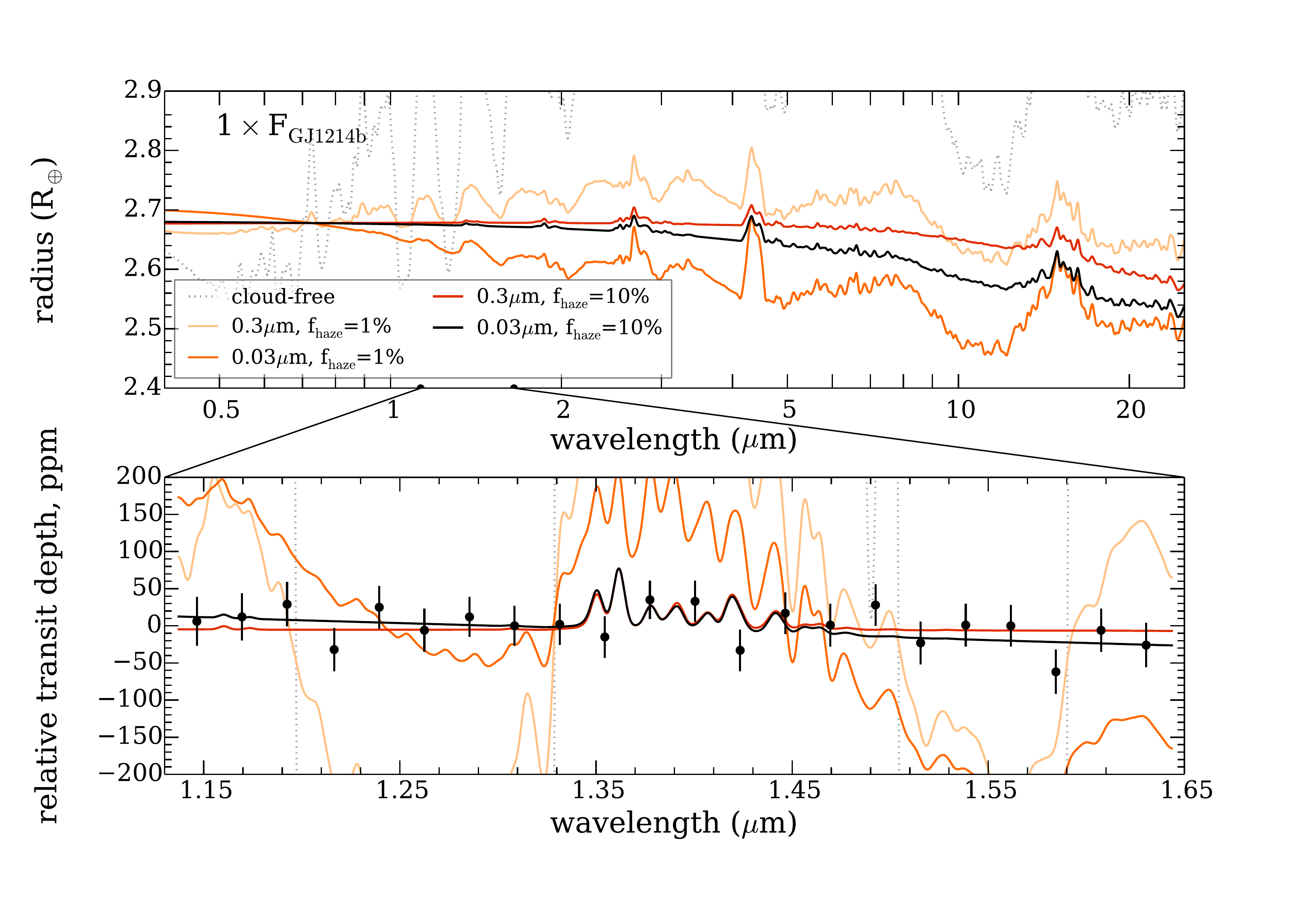}
\caption{Transmission spectra of models with photochemical hazes with two different mode particle radii (0.3 and 0.03 \micron) and \fhaze values (1 and 10\%). The top panel shows model planet radius from optical to mid-infrared wavelengths. The bottom panel shows the wavelength region (1.1--1.7 \micron) of the \ct{Kreidberg14} measurements. Note that the two models with \fhaze=10\% qualitatively match the flat spectrum.}
\label{trans_soots}
\end{figure*}

\subsection{Transmission Spectra}

\begin{figure*}[tb]
     \includegraphics[width=3.75in]{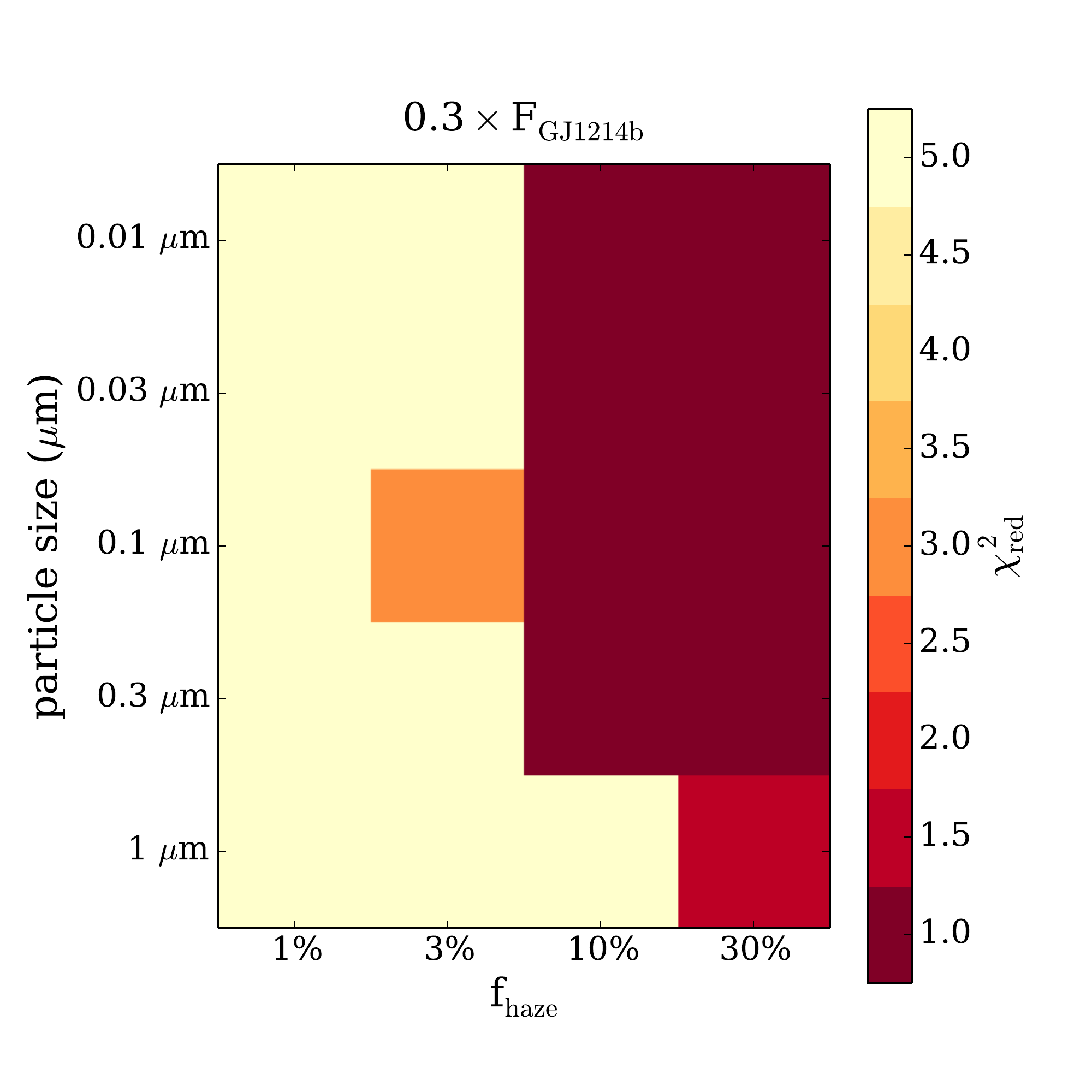}
     \includegraphics[width=3.75in]{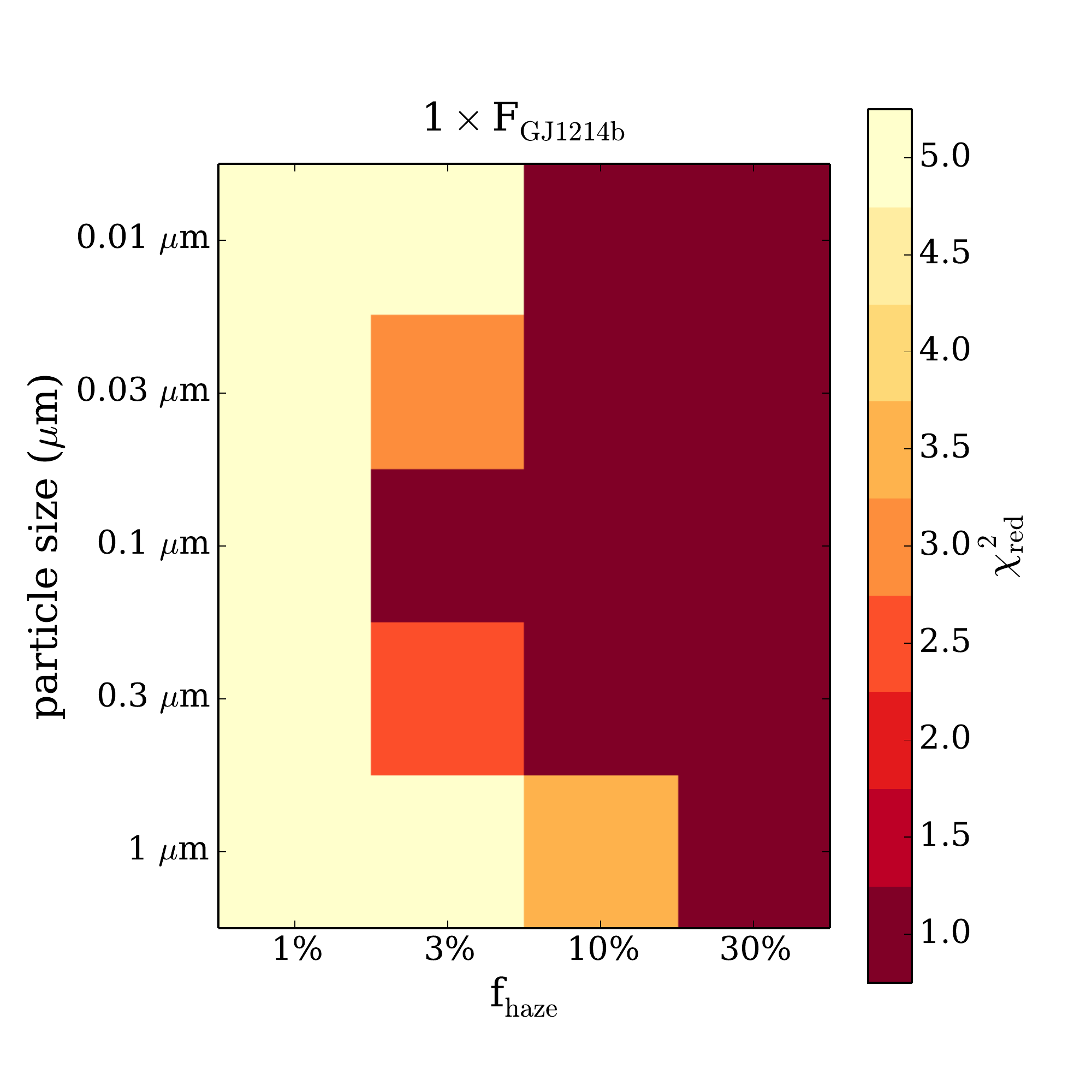}
     \includegraphics[width=3.75in]{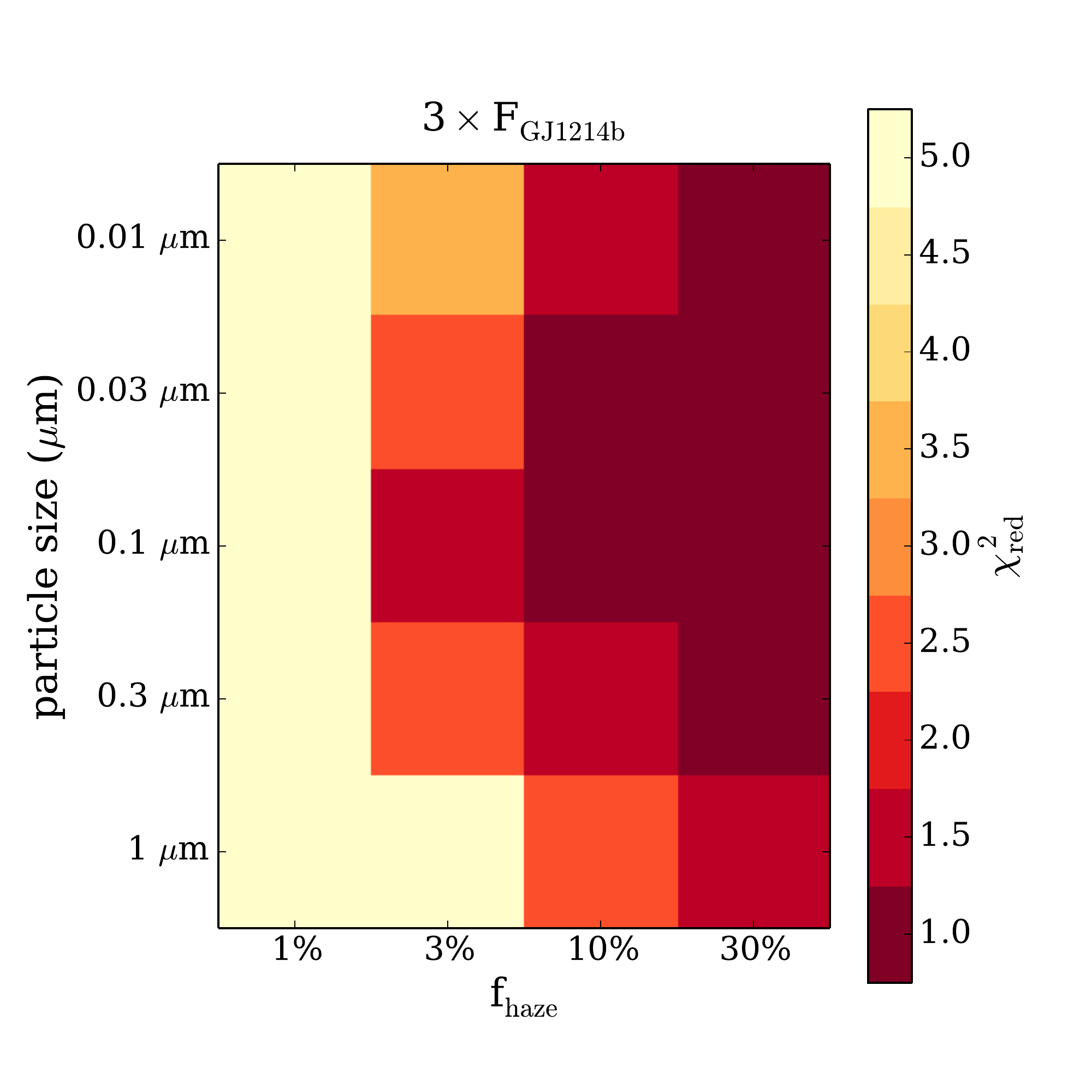}
     \includegraphics[width=3.75in]{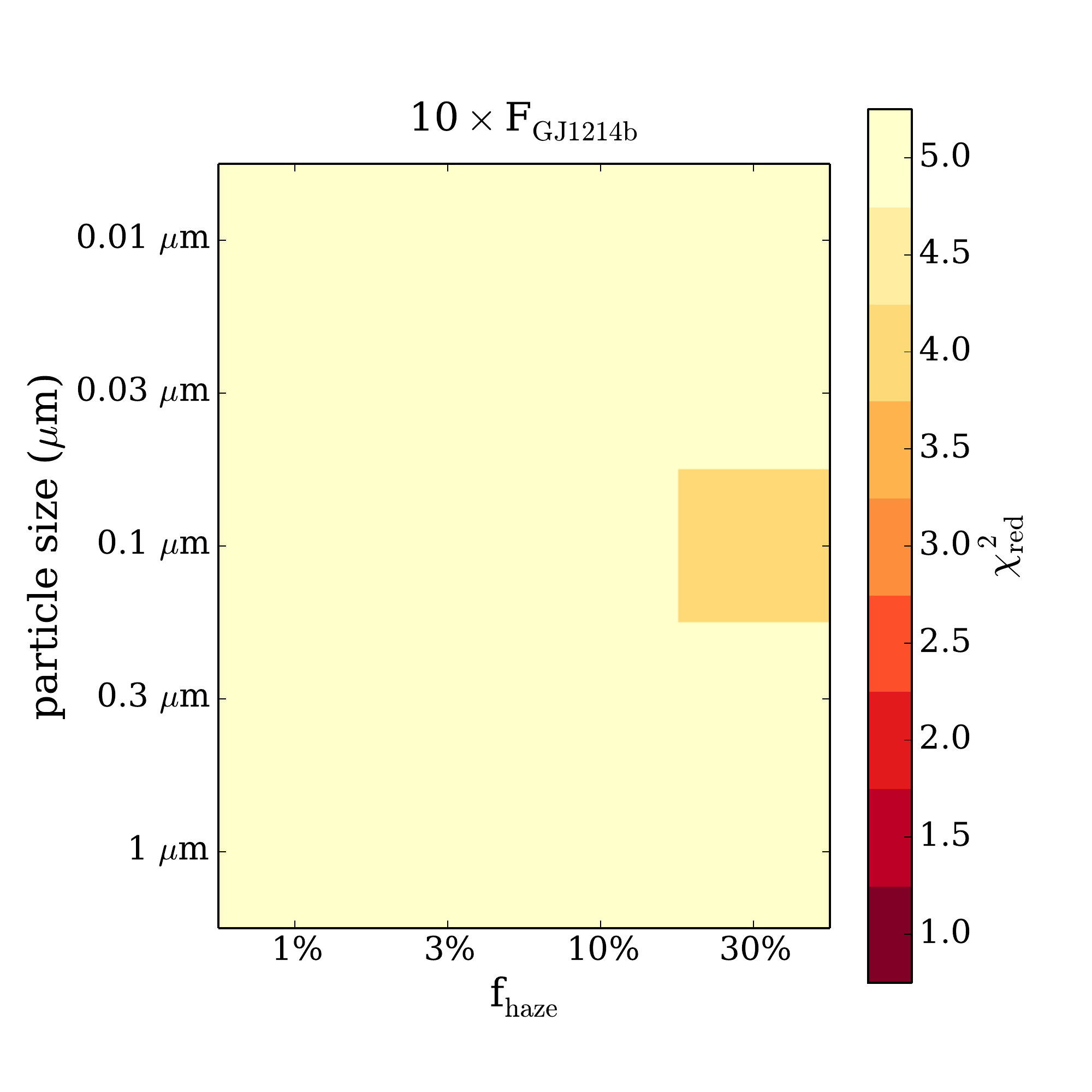}
 \caption{Chi-squared maps showing quality of fit to \ct{Kreidberg14} data for transmission spectra with photochemical hazes, with varied irradiation levels, mode particle sizes, and haze forming efficiency $f_{\rm haze}$. Starting in the top left panel, models with 0.3, 1, 3, and 10 $\times$ GJ 1214b's irradiation are shown. Dark red sections show acceptable fits (reduced $\chi^2$ close to 1.0). Note that a variety of models with \fhaze=10--30\% can generate a flat enough transmission spectrum to be consistent with the data, for models cooler than 10$\times$ GJ 1214b's irradiation (\teff$\sim$1100 K).}
\label{chisquare_map_soot}
\end{figure*}

Figure \ref{trans_soots} shows examples of model transmission spectra at GJ 1214b's incident flux. We summarize our results for a wider set of parameters in Figure \ref{chisquare_map_soot}. 

We find a few key results: 

\begin{enumerate}

\item A photochemical haze thick enough to flatten the near-infrared transmission spectrum only forms in models with 0.3--3$\times$ GJ 1214b's irradiation. Models at 10--30$\times$ GJ 1214b's irradiation are warmer and therefore have less methane (and more CO) resulting in overall less soot precursor material (see also Figures \ref{photochem_summed} and \ref{photochem_bar}). In addition, these warmer models have somewhat larger scale heights which means more soot material is needed to flatten the spectrum. 

\item Haze-forming efficiencies ($f_{haze}$) values of 10--30\% are necessary to flatten the spectrum for the assumed 50$\times$ solar composition. The value of $f_{haze}$ is essentially unconstrained in the literature due to the challenges of modeling all possible kinetics pathways to long chain hydrocarbons. 

\item Small particles ($r \le 0.1 \mu m$) have optical properties that cause them to absorb more efficiently at the shortest wavelengths. However, the hazes in this model become optically thick over a small range of height $z$, resulting in only a minor slope to the transmission spectrum even for particle sizes of 0.01 \micron. 

\end{enumerate}

 \begin{figure*}[tbh]
  \centering  \hspace{-1cm}
  	\begin{minipage}[t] {\textwidth}
  \includegraphics[width=7.7in]{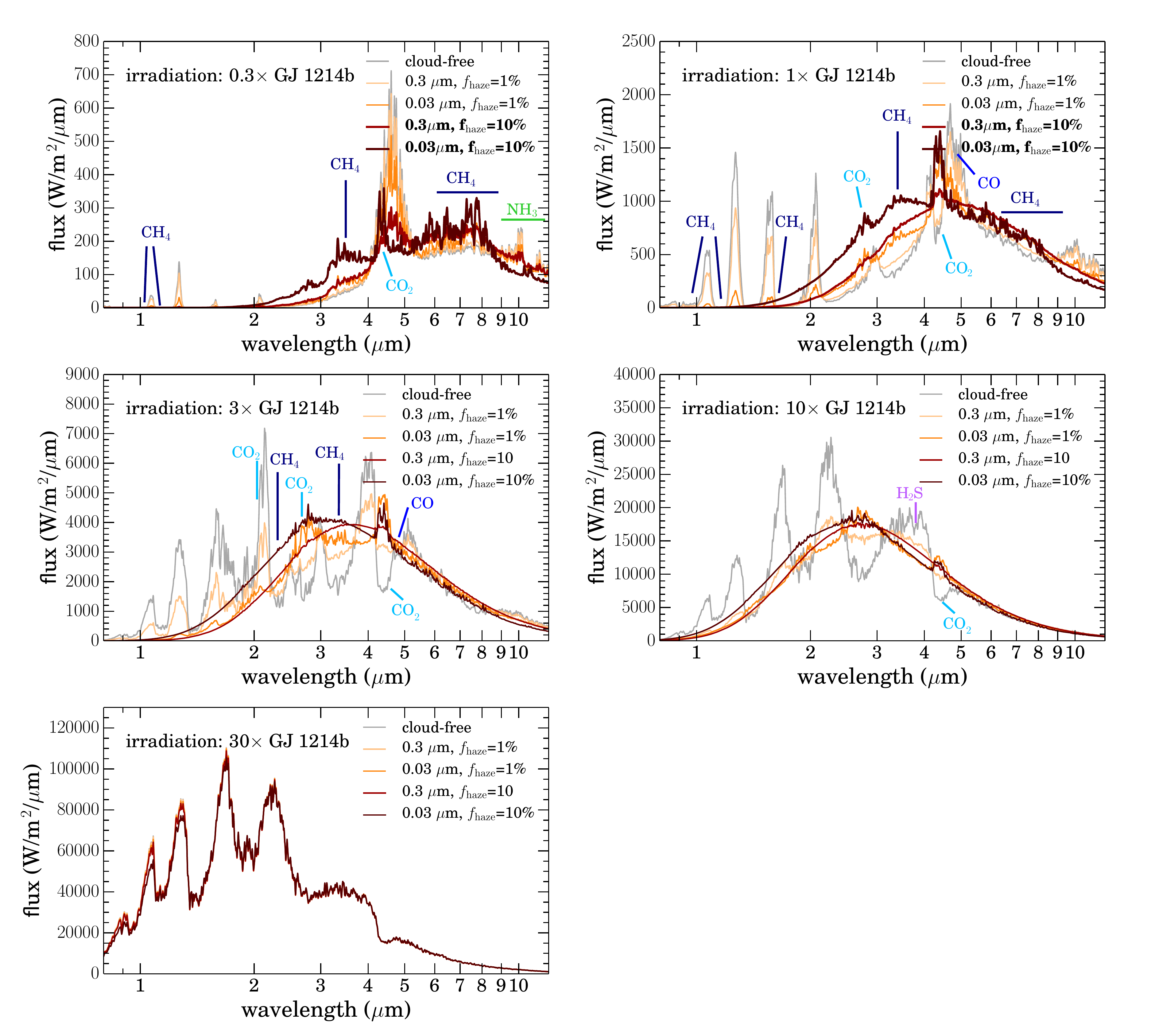}
	\end{minipage}
 \caption{Thermal emission spectra with photochemical haze. Each panel shows a different irradiation level. Cloud-free models are shown as gray lines; models with haze particle sizes of 0.03 and 0.3 \micron\ and \fhaze\ of 1 and 10\% are shown as colored lines, with hazier models in darker colors. The fonts in the captions are bolded if the transmission spectrum with those parameters fits the \ct{Kreidberg14} data.   }
\label{thermal_spec_soot}
\end{figure*}

\subsection{Thermal Emission Spectra}

Thermal emission spectra at each irradiation level are shown in Figure \ref{thermal_spec_soot}. The top right panel shows predictions for models with GJ 1214b's irradiation. At this temperature ($\sim$600 K), the spectrum shows absorption features from water, methane, and carbon monoxide. For thin hazes which do not flatten the transmission spectrum (\fhaze=1\%), the flux in the near-infrared peaks decreases, and the flux at absorption features, especially between 2 and 4 \micron, increases. These changes are due to increased cloud opacity and increased temperature of the P--T profile due to the absorption of stellar flux by particles in the upper atmosphere. For thick hazes, the heating in the upper atmosphere is large and causes a temperature inversion (see Section \ref{antigreen} and Figure \ref{pah_cond}). This causes some molecular features to be seen in \emph{emission} instead of absorption. Most prominent of these is CO$_2$, between 4 and 5 \micron; at GJ 1214b's irradiation, all hazes that flatten the transmission spectrum have CO$_2$ in emission. This feature is potentially observable with \emph{JWST} (see Section \ref{jwst}).  

At 3--10$\times$ GJ 1214b's irradiation, haze-free spectra have significant features, while hazy spectra have muted features and very little flux in the near-infrared. The emission bands are weaker at higher temperatures. At 30$\times$ GJ 1214b's irradiation, the hazes are optically thin and we see very little difference between the models. 

At 0.3 $\times$ GJ 1214b's irradiation, hazes decrease the flux in the near-infrared and in the 4--5 \micron\ window between water and methane features, and increase the flux between 2 and 4 \micron.

 \begin{figure*}[tbh]
     \includegraphics[width=7.5in]{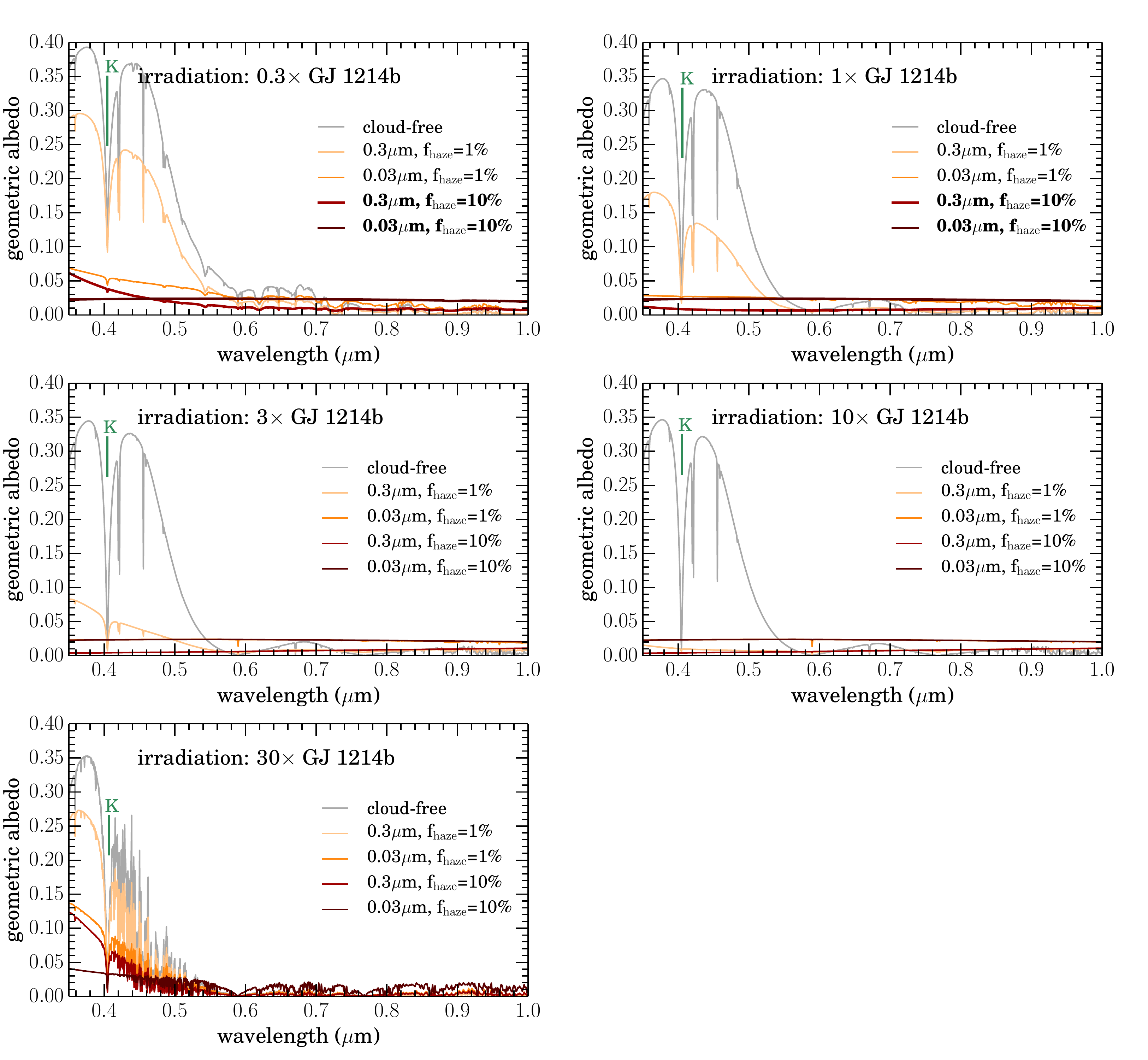}
 \caption{Albedo spectra with photochemical haze. Haze-free models are shown as gray lines; models with haze particle sizes of 0.03 and 0.3 \micron\ and \fhaze\ of 1 and 10\% are shown as colored lines, with hazier models in darker colors. The fonts in the captions are bolded if the transmission spectrum with those parameters fits the \ct{Kreidberg14} data. Note that the scale on these plots is different from the previous albedo spectra in Figures \ref{alb_specs_eq} and \ref{alb_specs_cold}.  }
\label{alb_spec_soot}
\end{figure*}

\subsection{Albedo Spectra}

Figure \ref{alb_spec_soot} shows albedo spectra for the same set of models as shown in Figure \ref{thermal_spec_soot}. 

Haze-free models are brightest between 0.3 and 0.55 \micron, with geometric albedos around 0.1 to 0.4, because Rayleigh scattering is most efficient at short wavelengths. At these short wavelengths, the hotter models have lower albedos than cooler models. From 0.6--1 \micron, the albedo spectra are quite faint, with geometric albedo $<$1--4\% because atoms and molecules absorb photons at higher altitudes than Rayleigh scattering reflects them. In particular, the pressure-broadened lines of the alkali metals absorb strongly at green and red optical wavelengths. 

Soot hazes cause dark reflected light spectra. This result is not surprising given the strongly absorbing optical properties of black soots and their high altitudes; the soots absorb visible light photons at higher altitudes would be scattered. Thin hazes decrease the reflected flux at all wavelengths, to 5--70\% of the haze-free albedos. For very thick hazes, the albedo becomes more uniformly dark, around 2\%. At longer wavelengths (0.6--1.0 \micron) the thick hazy model spectra are somewhat brighter than the very dark haze-free spectra; at wavelengths between 0.3--0.55 \micron, the thick hazy model spectra are darker than haze-free spectra, because the soot layer absorbs the visible photons at higher altitudes than photons would scatter by Rayleigh scattering.

 \begin{figure*}[tb]
\hspace{-10mm}
\center     \includegraphics[width=7.3in]{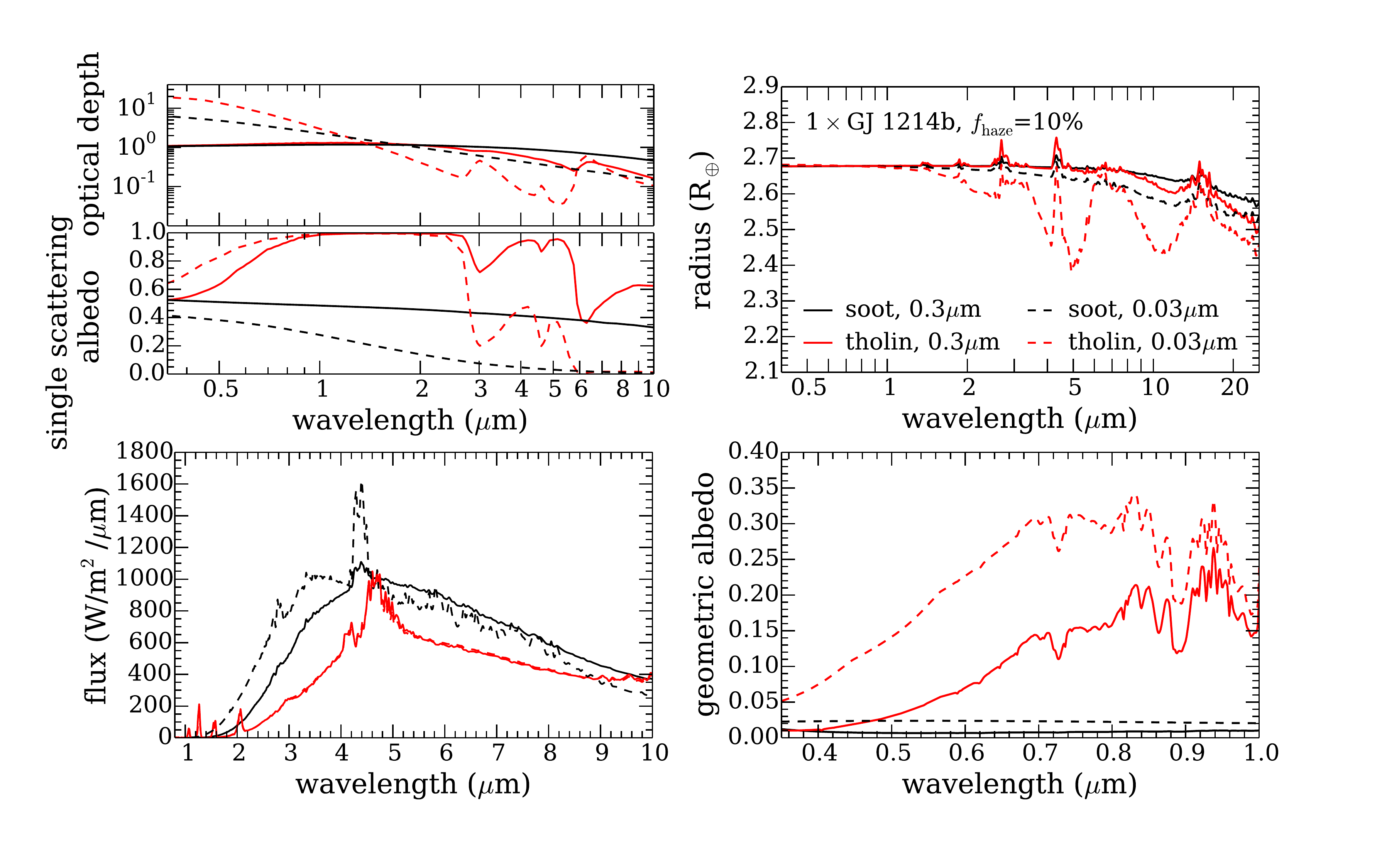}
 \caption{Effect of optical properties of photochemical haze on spectra. Each panel includes models with soot optical properties (black lines) and tholin optical properties (red lines) with two different particle sizes (0.3 and 0.03 \micron) as solid and dashed line styles. Top left: cloud optical depth and single scattering albedo; top right: transmission spectra; bottom left: thermal emission spectra; bottom right: geometric albedo spectra. }
\label{tholins}
\end{figure*}

\subsection{Effect of Optical Properties of Photochemical Haze}

An important assumption made in the nominal photochemical haze grid is that haze particles have the same optical properties as soots \cp{Hess98}. Real hazes likely have diverse optical properties that depend on the environment in which they form. Here, we show how key spectral features change if different optical properties are used. We change the optical constants to those of tholins, a material created in lab experiments to simulate hazes. Tholins are similar to the materials that form hazes in Titan's atmosphere, which form due to photochemistry at high altitudes. In a simulated transmission spectrum measured using a solar occultation with the Cassini spacecraft, this hydrocarbon haze produces a distinct slope from near- to mid-infrared wavelengths \cp{Robinson14b}. Titan's haze particles are made of fractal aggregates of large hydrocarbons \cp{McKay01}. We use tholin indices of refraction from the experimentally derived values in \ct{Khare84} and calculate absorption and scattering coefficients using Mie scattering assuming spherical particles. We hold all other properties constant---particle sizes and \fhaze, the haze number density, haze particle density---to isolate the effect of optical properties alone.
 
Figure \ref{tholins} summarizes these results. The top left panel shows the cloud properties at a single slice in the atmosphere, where the haze becomes optically thick in the near-infrared (1.5 \micron). The optical depth of the tholin haze depends strongly on wavelength for both particle sizes (higher optical depth at shorter wavelengths) and features are visible, especially for the smaller particle size. More dramatically, the single scattering albedo of the small tholin particles is high in the near-infrared ($\sim$1 from 1--2.5 \micron) with strong features in the optical and mid-infrared. In contrast, soot particles of both sizes have low, feature-poor single scattering albedo. 

 The top right panel shows examples of transmission spectra; note that small tholin particles absorb strongly at optical but not infrared wavelengths, unlike soots which absorb more uniformly across the infrared.  Because they are much less efficient infrared absorbers, none of the models with tholin optical properties adequately fit the \ct{Kreidberg14} data. 

The bottom left panel shows examples of thermal emission spectra. The most profound difference from the soot models is that, because the tholins absorb much less of the stellar irradiation, the upper atmospheres do not warm and form a temperature inversion. \footnote{This finding of course differs from Titan's actual atmosphere which does have a haze-caused temperature inversion \cp{McKay91}}. Without a temperature inversion, none of the emission features seen in the  spectra with soot haze are seen in spectra with tholin haze. In addition, more of the spectral features at near-infrared wavelengths are preserved. 

The bottom right panel shows albedo spectra. The first obvious change is that tholin hazes scatter much more efficiently, making the albedo spectra overall much brighter (15--30\% between 0.7 and 1 \micron). The tholin hazes absorb more efficiently at blue wavelengths (0.3--0.6 \micron), causing the spectrum to be darker at blue wavelengths and brighter at red wavelengths. Features from methane are easily visible around 0.9 \micron.

  \begin{figure}[t]
     \includegraphics[width=3.7in]{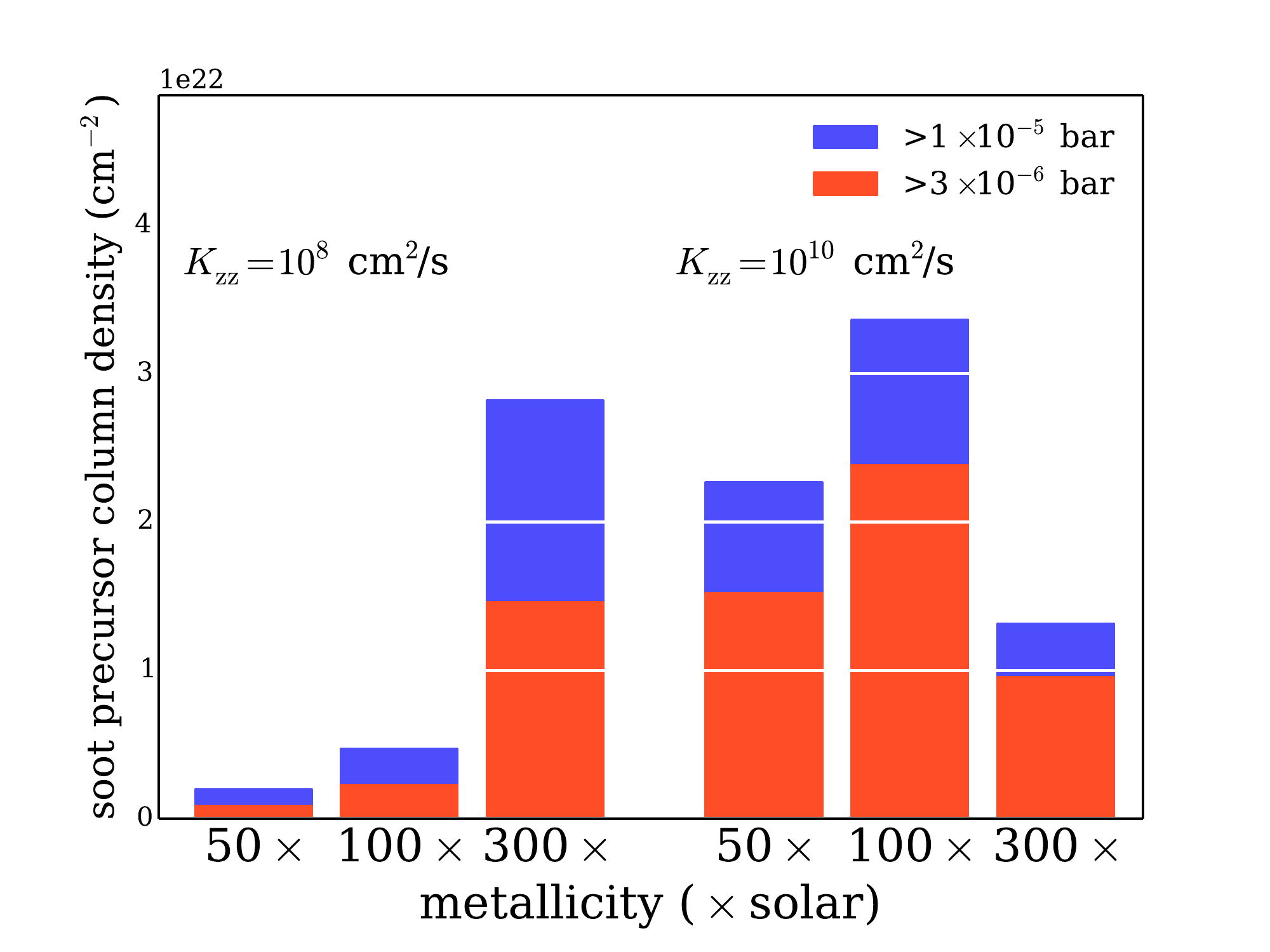}
 \caption{Effect of \kzz\ and metallicity on column density of soot precursors, at incident flux of GJ 1214b. Photochemical models with \kzz=10$^8$ cm$^2$/s are on the left and \kzz of10$^{10}$ cm$^2$/s are on the right. At lower \kzz, the column densities of high altitude soot precursors increase substantially with increased metallicity. At higher \kzz, there is a peak at 100$\times$ solar metallicity and no clear trend. }
\label{phot_highMH}
\end{figure}

\subsection{Photochemistry At Higher Metallicities}

All of the hazy models presented here assume compositions of 50$\times$ solar metallicity; however, the metallicities of low mass, low density planets may be higher (see Section \ref{highmetlikely}).  There are several competing effects that control the formation of hazes in higher metallicity atmospheres and the amplitude of features in their transmission spectra. Higher metallicity atmospheres ($>50\times$ solar) have higher mean molecular weights and therefore smaller scale heights, reducing the amplitude of features. The amount of carbon available increases (by definition) uniformly at higher metallicities. However, the abundance of soot precursors available to form hazes does not necessarily follow, due to the complex interactions of kinetic pathways to make and destroy soot precursors. 

To create soot precursors, an atmosphere must be methane-rich. High metallicity tends to favor the production of CO and CO$_2$ over CH$_4$, which can potentially inhibit soot precursor production. In a methane-dominated atmosphere, vigorous mixing (high \kzz) increases soot precursor production (see Figures \ref{photochem_summed} and \ref{photochem_bar}). However, vigorous mixing can also increase the abundance of CO and CO$_2$ and decrease the abundance of CH$_4$, which decreases soot precursor production. 

Examples of high metallicity models are shown in Figure \ref{phot_highMH}. We find that for less vigorous mixing (\kzz=10$^8$ cm$^2$/s), the column density of soot precursor formed at high altitudes increases with increased metallicity, at a rate higher than would be predicted by the increase in carbon abundance alone. In contrast, with more vigorous mixing (\kzz=10$^{10}$ cm$^2$/s) the column density of soot precursor formed is largest at 100$\times$ solar. 

More work should be done in the future to fully understand the differences in kinetics pathways at high metallicity, but, generally, we find that planets with a variety of metallicities can have similarly rich photochemistry that likely allows for the formation of hazes.

\section{Discussion} \label{discussion}

\subsection{High Metallicity Super Earth Atmospheres}\label{highmetlikely}

There are several lines of reasoning that suggest that small, gas-rich planets may have high metallicities. 

The first is purely empirical. In the solar system, there is a power law relationship between planet mass and metallicity, with lower mass planets being significantly more enhanced in heavy elements. Based on carbon abundance derived from methane, Jupiter (318 \me) is 3.3--5.5$\times$, Saturn (95 \me) is 9.5--10.3$\times$, Uranus (14.5 \me) is 71--100$\times$, and Neptune (17 \me) is 67--111$\times$ solar metallicity \cp{ Wong04, Fletcher09, Karkoschka11, Sromovsky11}. \ct{Kreidberg14b} extend this comparison to a more massive exoplanet, WASP-43b, which has a mass of 2\mj\ and a metallicity (based on the measured water abundance) of 0.4--3.5 $\times$ solar. 

Extrapolating this power law to GJ 1214b's mass ($\sim$6\me) results in a predicted metallicity of 200--300$\times$ solar. Of course, nature need not continue to follow this particular power law if, for example, the formation mechanism for extrasolar small planets differs significantly from the gas and ice giants in our own solar system, but this line of reasoning provides a testable prediction. 

The other line of reasoning is based on population synthesis models of super Earths. \ct{Fortney13} show that, based on models that follow the accretion of gas and planetesimals to form planets, objects in the super Earth mass range may have a wide diversity of envelope enrichments. They predict that a portion of the population will have highly enriched atmospheres of several hundreds of times solar composition (see Figure 5 from \ct{Fortney13}). 

Together these lines of evidence show that high metallicities may be quite common, and that a measurement of atmosphere enrichment for a planet smaller than Uranus would be valuable for our understanding of planet formation.

\subsection{Is \fsed=0.01 Reasonable?}\label{lowfsedlikely}

For a cloudy planet to have a flat transmission spectrum, the atmosphere must both have high metallicity and inefficient cloud sedimentation (\fsed$\ll$1). This low inferred \fsed\ is much less than the inferred \fsed\ for brown dwarfs (\fsed$\approx$1--5). 

However, clouds flattening GJ 1214b's spectrum need not behave the same as the deep convective iron and silicate clouds of brown dwarfs. In fact, we might expect them to behave more like stratospheric clouds on Earth. When parameterized with this model, terrestrial stratocumulus clouds have \fsed$<1$ at the top of the cloud, with increasing \fsed\ with distance below the cloud top. Clouds studied over the North Sea, for example, have been measured to have \fsed$\sim$0.2 \cp{AM01}. It is possible that GJ 1214b differs enough in circulation patterns from Earth, as a tidally locked planet around an M dwarf, that clouds in the upper atmosphere could be more vigorously lofted to create even lower \fsed\ clouds. 

Further study is needed to determine whether these values are reasonable (e.g. 3D circulation models with radiatively-interacting cloud tracer particles would inform us about where the clouds are likely to form). 
  
   \begin{figure}[t]
     \includegraphics[width=3.7in]{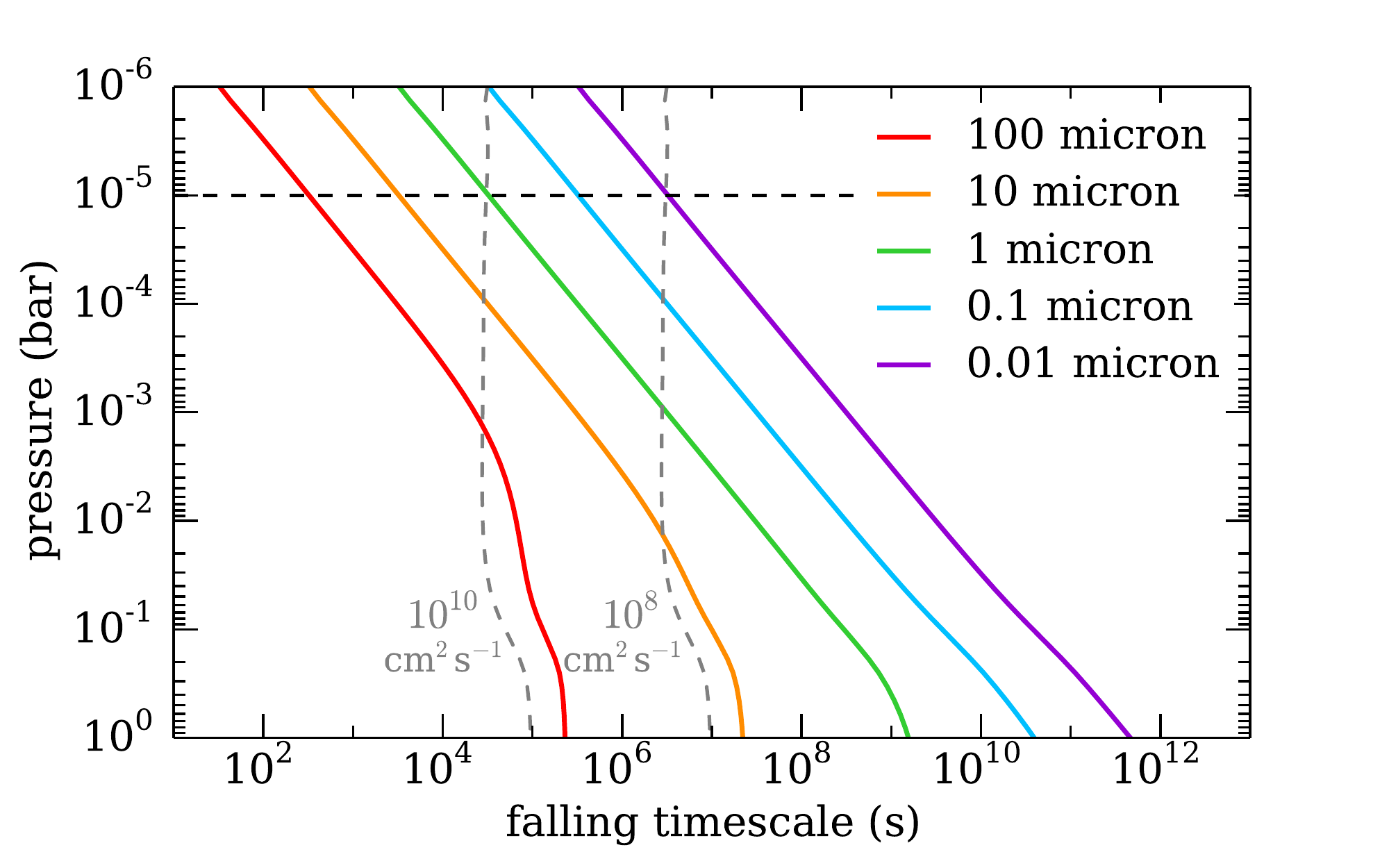}
 \caption{Cloud particle falling timescales. The dashed horizontal line is at 10$^{-5}$ bar, the approximate height of GJ 1214b's haze. Solid lines show the timescale for particles to fall one pressure scale height as a function of particle size. The dashed vertical lines show the pressure scale height divided by constant \kzz\ (10$^8$ and 10$^{10}$ cm$^2$s$^{-1}$), giving the ``lofting timescale" for that \kzz.  }
\label{lofting_particles}
\end{figure}
 
 \subsection{Vertical Mixing to Loft Small Particles}
 
 Our results for photochemical hazes suggest that they may provide a viable way to flatten the transmission spectra of small planets. However, the models were not run with a self-consistent cloud model that governs how fast particles can sink out of the atmosphere. In particular, can particles with sizes from 0.01--0.3 \micron, which allow us to fit the data, stay lofted for long enough timescales for new particles to form? 
 
 We do not attempt to address these questions here, without a complete model for cloud formation in a planetary atmosphere, nor a model for 3D atmospheric circulation, both of which would be necessary to address this question. We can however show that, for our assumed vertical mixing values in the photochemical model (\kzz=10$^{10}$ cm$^2$s$^{-1}$), which are based on upper limits from circulation models (\ct{Kataria14} and T. Kataria, private communication), mixing should be vigorous enough to loft $\sim$1 \micron\ particles. 
 
 In Figure \ref{lofting_particles}, we show the timescale for a cloud particle to fall one pressure scale height ($H/v_{\rm fall}$) where $H$ is the scale height and $v_{\rm fall}$ is the particle falling velocity. We calculate falling velocities assuming viscous flow, following the approach of \ct{AM01} (their Appendix B).   We also show lines that represent constant \kzz\ of 10$^{8}$ cm$^2$s$^{-1}$ and 10$^{10}$ cm$^2$s$^{-1}$, which were the values used in the photochemical models. 
  
 We find that for particles smaller than 1 \micron, the falling timescale is longer than the lofting timescale assuming \kzz=10$^{10}$ cm$^2$s$^{-1}$. Given these conditions, it should therefore be possible to have particles of this size in the upper regions of GJ 1214b's atmosphere. However, if the mixing is less vigorous, it will be significantly harder to keep particles in the size range from 0.01--1 \micron\ lofted at 10$^{-5}$ bar.

 \subsection{Need for Laboratory Studies at Super Earth Conditions} 
  
  One path forward to understand photochemical hazes is the same that has been used for decades to study Titan's complex atmospheric chemistry: laboratory measurements. The conditions present in super Earths like GJ 1214b, including the moderately high temperature ($\sim$600 K) and the H$_2$-rich composition, are quite different from that of any solar system planets or moons, and therefore require new laboratory studies. 
  
  Lab experiments are crucial because theoretical modeling of full chemical kinetic pathways from 2-carbon hydrocarbons to complex PAHs and long-chain hydrocarbons poses a huge challenge. The information provided by laboratory measurements would provide empirical constrains on these reactions. For example, we could determine whether reactions necessary to create condensible hydrocarbons do indeed proceed at low pressures in a GJ 1214b-like atmosphere, and whether, like on Titan's these hydrocarbons form with the help of ion chemistry \cp{Lavvas11}. The types of condensed materials could be predicted and their optical properties would allow us to make predictions for future observations. The concentrations of other gases formed in the chemical reactions could be determined and testable predictions could be made. In addition, lab experiments could allow us to make predictions, beyond the predictions we make here, about which conditions create the most obscuring haze material, allowing us to better target planets.  
  
\subsection{Planning Future Observations of Super Earths} \label{jwst}

The \emph{Kepler} results demonstrably show that super Earths are incredibly common. To understand planets as a population, we must be able to measure properties of super Earths. The flat transmission spectra of super Earths that have been observed over the last few years have shown that this is not as easy a task as originally perceived \cp[e.g.,][]{Miller-Ricci09}. We suggest several directions that may allow us to move forward to understand the compositions of super Earth atmospheres. 

\subsubsection{Transmission Spectra of Hotter Targets}

One avenue for advancement is to observe warmer super Earth targets. If photochemical hazes are indeed obscuring the transmission spectra of cool targets such as GJ 1214b, these hazes, according to our models, should decrease in abundance significantly between 3 and 10$\times$ GJ 1214b's irradiation (around $\sim$1000 K), at the transition between CO and CH$_4$ dominated compositions (see Figures \ref{photochem_summed} and \ref{photochem_bar}). We note that we do not consider hazes derived from other elements such as sulfur, which may exist at warmer temperatures \cp{Zahnle09b}. 

This idea has also been discussed in \ct{Fortney13} (see their Figure 6), and one of the best targets, since it is $\sim$2000 K and around a bright star, is 55 Cnc e. A handful of \kepler\ planets are also >1000-1100 K, but orbit faint stars that make the observations challenging. In addition, many small planets in this temperature range may have experienced significant mass loss \cp[][their Figure 1]{Lopez12, Lopez13, Fortney13}. The current K2 mission (using the repurposed \kepler\ telescope) \cp{Howell14} and upcoming Transiting Exoplanet Survey Satellite (TESS) mission \cp{Ricker14} may reveal additional hot super Earths around the stars they target, which are on average closer and brighter than the \kepler\ targets. 

Mapping out the parts of parameter space with flat transmission spectra will provide information about the types of clouds and hazes that exist in these atmospheres. Temperature (incident flux) is the most important parameter that likely controls clouds and hazes; unfortunately most of the targets observed so far have been in the same 600--900 K range that we predict to have significant methane-derived photochemical hazes.

\subsubsection{Thermal Emission Spectra with JWST}

Looking to the future, one path that will be opened with the launch of the James Webb Space Telescope (\jwst) will be observing the thermal emission spectra of warm and hot super Earths. These will be challenging measurements that will likely take several secondary eclipses to achieve the necessary signal-to-noise to detect features (Greene et al., in prep.). 

Several instruments will be capable of observing secondary eclipses of super Earths. In the near-infrared, both the Near-Infrared Camera (NIRCam) and Near-InfraRed Imager and Slitless Spectrograph (NIRISS) will be able to observe transits and eclipses. In particular, NIRCam has a grism mode that will be capable of 2.4--5 \micron\ R$\sim$2000 slitless spectroscopy. It uses a slitless grism that is sensitive to sky background across a large field, which is optimized for the precision photometry and stability needed to make these observations of exoplanets. NIRISS has a single object slitless spectroscopy mode with wavelength coverage from 0.6--2.5 \micron\, spectral resolution of $\sim$700, and optimized for spectroscopy of transiting planets. Lastly, The Near-Infrared Spectrograph (NIRSpec), offers slit spectroscopy in the 0.6 to 5.0~$\mu$m wavelength range with a wide variety spectral resolutions (30 < R < 3500), which may be particularly useful for targeted observations of specific spectral features.  

For longer wavelengths, the Mid-Infrared Instrument (MIRI) will be capable of low (R$\sim$100) resolution spectroscopy from 5--14 \micron\ and moderate resolution (R$\sim$3000) spectroscopy from 5--28.3 \micron. It is the only \jwst\ instrument that will observe wavelengths longer than 5 \micron\ and will be 50 times more sensitive than the Spitzer Space Telescope.

\begin{figure}[t]
     \hspace{-5mm}
     \includegraphics[width=3.9in]{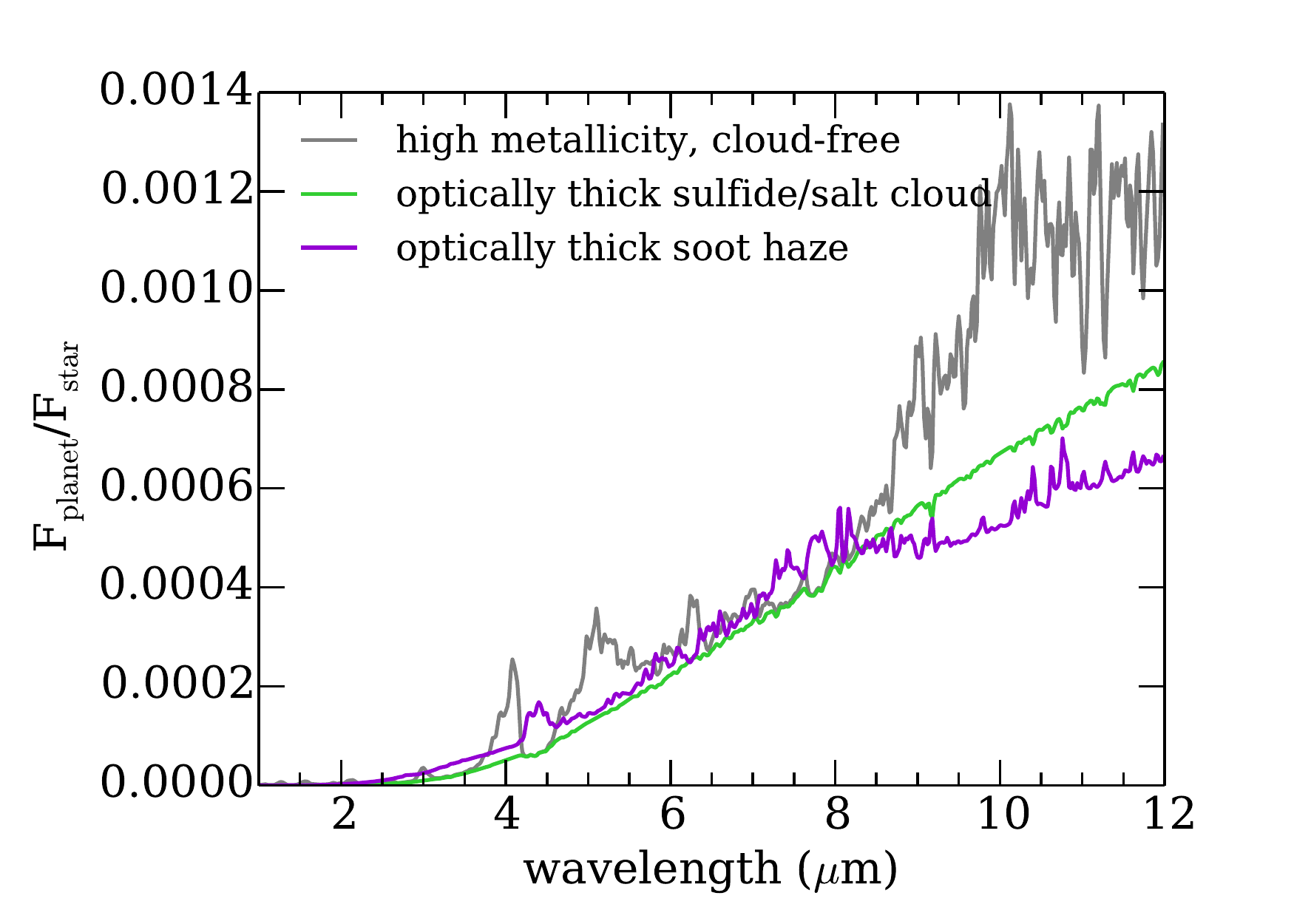}
 \caption{Planet star flux ratio of cloud-free, cloudy, and hazy GJ 1214b analogs. Thermal emission spectra are divided by a blackbody representing the GJ 1214b host star. Models are smoothed to R$\sim$200. All models are at GJ 1214b's incident flux. Cloud-free and cloudy model are 1000$\times$ solar metallicity, and the cloudy model has cloud parameter \fsed=0.01 (\nas, KCl, and ZnS clouds). The hazy model has mode particle size of 0.03 \micron\ and \fhaze=10\%.  } 
\label{jwst_plot}
\end{figure}

Figure \ref{jwst_plot} shows the planet-star flux ratio (i.e. the depth of secondary eclipse) for three different representative models. Given high signal-to-noise observations across a wide wavelength range, it should be possible to determine the differences between these models. However, piecing together an infrared spectrum will be an expensive endeavor that requires multiple observations, each taking many hours. As a community, targets for this treatment must be carefully considered. 

Of concern is that many models with thick clouds (that match the transmission spectrum observations) have spectra that appear nearly identical to blackbodies. If these models indeed represent reality, thermal emission will not allow us to determine the compositions of gases in the planetary atmosphere. However, the models that include optically thick photochemical hazes in the upper atmosphere have strong temperature inversions that create observable emission bands in the mid-infrared. Discovering a spectrum like this would strongly indicate that hazes are indeed the cause of flat transmission spectra; constraining the strength of the temperature inversion would allow us to constrain the optical properties of the hazes, since this inversion indicates that the particles are strong optical and weaker infrared absorbers.  

\subsubsection{Albedo Spectra from Space-based Coronagraph}

Further in the future, a space-based mission with a coronagraph, such as the \emph{WFIRST-AFTA} mission, will allow us to measure the reflected light from old, giant planets, just as we have observed the solar system planets for centuries. Current predictions for the performance of the \emph{WFIRST-AFTA} coronagraph suggest that for favorable configurations, super Earths and small Neptunes may also be viable targets \cp{Spergel15}. These objects will be easily observable with a larger space-based telescope designed to be capable of characterizing habitable-zone Earth-like planets (e.g. Advanced Technology Large-Aperture Space Telescope (ATLAST), Terrestrial Planet Finder (TPF), or High Definition Space Telescope (HDST)).

\begin{figure}[t]
     \includegraphics[width=3.7in]{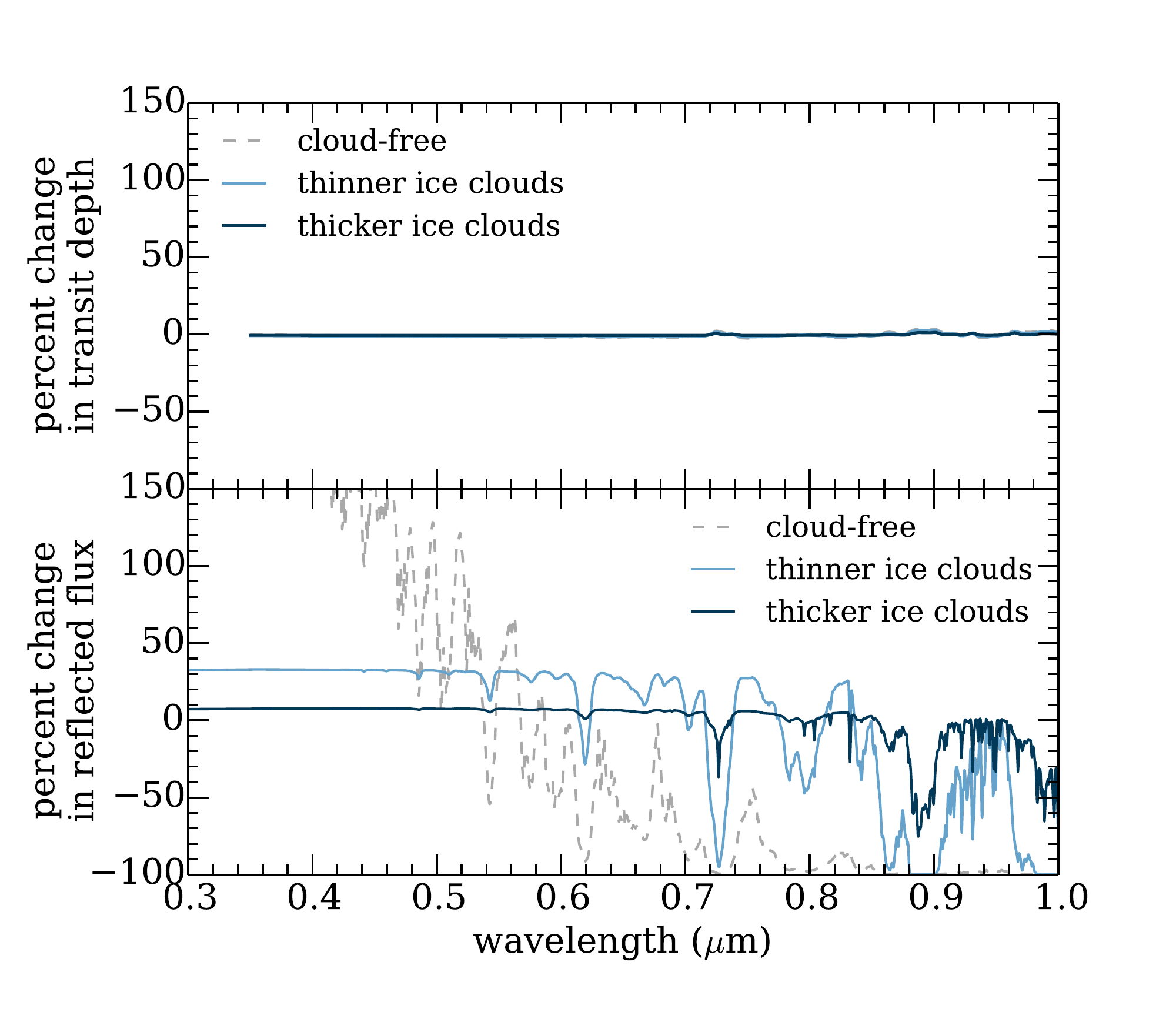}
 \caption{Relative amplitude of measurement compared to mean for transmission spectra (top) and reflected light spectra (bottom) for a planet with 1\% GJ 1214b's incident flux, 50$\times$ solar composition, and \fsed=1 and 0.1 for the thinner and thicker clouds respectively. The percent change in transit depth in transmission is very small, regardless of the molecules present (the cloud-free and thinner clouds lines plot are covered by the thicker clouds line). The percent change in reflected light will be up to several hundred percent, with the planet disappearing at wavelengths of very strong absorption features and becoming very bright at wavelengths with efficient scattering. As a caveat, note that the precision achievable during a transmission spectrum observation is much higher than the precision achievable in a reflected light measurement.  }
\label{dual_fig}
\end{figure}

Figure \ref{dual_fig} shows the relative sizes of the features we might observe in reflected light compared to in transmission. In transmission, the radius of the planet changes by tiny amounts due to absorption by gases through the limb of the planet's atmosphere. The observable---the transit depth---changes by only a few percent. In contrast, in reflected light, the size of features may be large. Within deep absorption bands, the planet may disappear nearly completely (100\% change in reflected flux) compared to its average flux. At brighter-than-average wavelengths, it can be 100--200\% brighter. 

Reflected light from cold planets will be a rich source of information. Cold planets likely have thick layers of volatile clouds such as water and ammonia. Unlike in transmission spectra, where clouds tend only to damp spectral features, in reflected light, clouds actually make many features larger. Without clouds, only blue wavelengths have efficient scattering (from Rayleigh scattering by H$_2$ gas). At longer wavelengths, very little starlight is scattered, and the planet just appears uniformly dark. With clouds, especially volatile clouds which scatter very efficiently, light will scatter from the cloud layers. If layers above the cloud have gases with strong absorption bands, wavelengths within those bands will appear dark. The depth of the cloud, the composition of gas above it, and the strength of the band itself all affect the size of these molecular features. By measuring the depths of several features, we can therefore extract these pieces of information. Solar system scientists have been applying these techniques for decades, and we can draw on this knowledge base as we observe exoplanets in reflected light.

\subsubsection{High Resolution Spectra from Large Ground-based Telescopes}

Another fruitful path forward to measure the compositions of hazy planets may be to observe them at very high spectral resolution ($R\ge10^5$). Within the cores of spectral lines, the opacity is significantly higher than the average opacity across a molecular band. This means that, even with an obscuring haze, features may still be visible from absorption at the cores of these lines from the tenuous atmosphere above the haze \cp{Kempton14}. In the next decades, these observations may be possible using the thirty meter class telescopes currently planned, such as the Thirty Meter Telescope (TMT), Giant Magellan Telescope (GMT), and E-ELT (European Extremely Large Telescope).

\section{Conclusions} \label{conclusion}

We have presented models of low mass, low density planets to explore the effect of clouds and hazes which are known to be present in super Earth atmospheres such as GJ 1214b. The grids of models are GJ 1214b analogs in their gravity, radius, and host star, and span a wide range of incident flux, metallicity and cloud properties. Key insights of this study include: 

\begin{enumerate}
\item For cloudy atmospheres to have featureless transmission spectra, they must have both very high metallicities ($\sim$1000$\times$ solar) and very inefficient cloud sedimentation compared to other clouds (\fsed$\sim$0.01). These characteristics seem possible but not the most probable scenario. 

\item Photochemical hazes likely form at high altitudes in planets like GJ 1214b. Assuming 50$\times$ solar composition, a variety of different haze particle sizes ($<$1 \micron) and haze forming efficiencies (\fhaze$\ge$10\%) can create featureless transmission spectra over a wide range in wavelength. 

\item Methane-derived photochemical hazes will not form in planets with \teff$\gtrsim$1000 K. Determining the prevalence of small planets with featureless transmission spectra over a range of incident flux will test this prediction. 

\item Thermal emission spectra of these planets will be possible to attain with dedicated \jwst\ time, and cloudy and hazy models may have distinct thermal emission. Cloudy thermal emission spectra have muted features and blackbody-like spectra. Photochemical hazes, depending on their optical properties, may cause mid-infrared emission features due to haze-caused temperature inversions.

\item Analysis of reflected light can distinguish between cloudy and hazy planets. Salt and sulfide clouds cause brighter albedos and potentially have features from optical properties of the clouds themselves such as ZnS at 0.53 \micron. Albedos of soot-rich planets will be very dark ($A_g$$\sim$2\%). 

\item Spectra of cold planets ($\sim$200 K) with ice clouds, potentially accessible to space-based coronagraphic telescopes like \emph{WFIRST-AFTA}, will have high albedos and information-dense molecular features, and may be a key population to study to measure super Earth compositions. 
\end{enumerate}

Despite the challenges presented by clouds and hazes in super Earth atmospheres, there are many paths forward for understanding super Earths in the next decades. At the present, we predict that observing warmer targets (>1000 K) with \emph{HST} will allow us to measure spectral features, because these objects should have a much less significant photochemical haze. Regardless of whether this prediction is correct, these measurements will allow us to determine which clouds and hazes are important. In the next decade, \jwst\ will measure thermal emission spectra of these small planets for the first time, and potentially place constraints on the optical properties of an optically thick haze. In future decades, observing reflected light from cold planets will be a leap in information content in our spectra and will allow us to better understand this population of super Earths.

\vspace{1in}

\begin{minipage}[h] {0.45\textwidth}

\acknowledgements 
\vspace{1in}
We thank the anonymous referee for their exceptionally helpful report which improved the manuscript. We also acknowledge the work to reformat our opacity database for the new radiative transfer code by high school students Anjini Karthik and Matthew Huang during Summer 2013. CVM acknowledges HST Theory Grant HST-AR-13918.002-A. JJF acknowledges Hubble grants HST-GO-13501.06-A and HST-GO-13665.004-A and NSF grant AST-1312545. MSM acknowledges support of the NASA Origins program. MRL acknowledges support provided by NASA through Hubble Fellowship grant \#51362 awarded by the Space Telescope Science Institute, which is operated by the Association of Universities for Research in Astronomy, In., for NASA, under the contract NAS 5-26555.
\end{minipage}

\vspace{0.5in}

\bibliographystyle{apj}

\clearpage

\appendix \label{append}

To model the thermal emission emerging from atmospheres of arbitrary composition, we developed a flexible new tool using the C version of the open-source radiative transfer code \texttt{disort} \cp{Stamnes88, Buras11}.  The code \texttt{disort} is a numerical implementation of the discrete-ordinate method for radiative transfer and is a powerful tool for monochromatic (unpolarized) radiative transfer, including absorption, emission, and scattering, in non-isothermal, vertically inhomogeneous media. It has been used for a variety of atmospheric studies in Earth's atmosphere and beyond, and here we apply it in a way that is applicable to self-luminous or irradiated exoplanets and brown dwarfs. 

In this calculation, \texttt{disort} takes as inputs arrays of optical depth ($\tau$), single scattering albedo ($\omega$), asymmetry parameter ($g$), and temperature ($T$). The flux and intensities are returned for a given wavenumber. For multiple scattering media, several treatments of the phase function are possible within \texttt{disort}'s framework; we implement the Henyey-Greenstein phase function. 

The bulk of the new calculations are written in the Python programming language. The radiative transfer scheme \texttt{disort} is in C and is called as a shared library from the main Python code. 

In order to calculate the emergent spectrum, we calculate $\tau$, $\omega$, $g$, and $T$ using the outputs of our 1D radiative--convective equilibrium code. We calculate spectra using models with 60 layers (though arbitrary numbers of layers are trivial to implement) and specify the temperatures at the 61 intersections between layers. Here we use molecular abundances calculated assuming equilibrium chemistry (though arbitrary compositions are also trivial to implement). 

Our opacity database is based on \ct{Freedman08} with significant updates described in \ct{Freedman14}, including methane \cp{Yurchenko14}, phosphine \cp{SousaSilva15}, and carbon dioxide \cp{Huang13,Huang14} . We include line lists of 17 molecules: H$_2$, He, CO$_2$, H$_2$O, CH$_4$, CO, NH$_3$, PH$_3$, H$_2$S, Na, K, TiO, VO, FeH, CrH, Rb, and Cs. It is very easy to add additional molecules to the model if we have line lists for their opacities. We include collision-induced opacity of H$_2$--H$_2$, H$_2$--He, H$_2$--H, and H$_2$--CH$_4$ using \ct{Richard12}. Rayleigh scattering is calculated for H$_2$, He, and CH$_4$ and is assumed to be isotropic \cp{Rages91}. We calculate line lists at 1060 pressure--temperature pairs from 10$^{-6}$ bar to 300 bar and 75 K to 4000 K at 10$\times$ the desired resolution (in this case, 1 cm$^{-1}$ resolution for a final resolution of 10 cm$^{-1}$). We interpolate the opacities bilinearly in $\log(P)$ and $\log(T)$ space to the pressures and temperatures of the P--T profile. We use Mie scattering (within the \ct{AM01} cloud code described in the Methods section) to calculate the single scattering albedo $\omega$ and asymmetry parameter $g$ of the clouds for each layer at each wavenumber. We sum all opacity sources, multiplying by the appropriate abundances, and convert opacities into optical depth $\tau$ by assuming hydrostatic equilibrium,
\begin{equation}
\tau=\dfrac{\Delta P}{\mu g}\sigma
\end{equation}
where $\Delta P$ is the change is pressure across a layer, $\mu$ is the mean molecular weight, $g$ is the gravity, and $\sigma$ is the opacity (cm$^2$ per atom or molecule).

Using the calculated values of $\tau$, $\omega$, $g$, and $T$, we call \texttt{disort} to calculate the flux at each wavenumber. 

 \begin{figure}[h]
     \centering \includegraphics[width=5in]{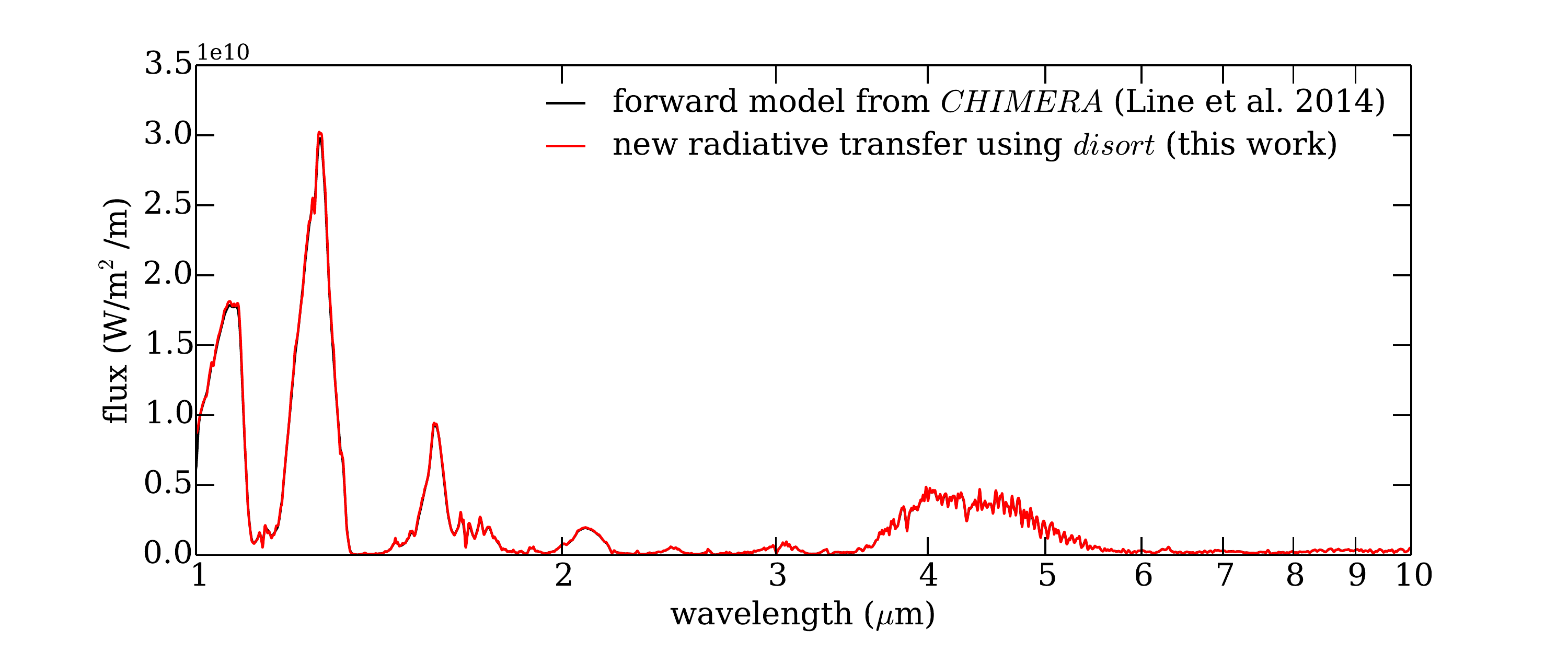}
 \caption{Comparison between radiative transfer methods at \teff=700 K, g=3000 m s$^{-2}$, cloud-free. Our test model from this work is shown in red; a spectrum using identical inputs (line lists, abundances, pressure, temperature) calculated using \texttt{CHIMERA} is shown in black. Note the excellent agreement at all wavelengths. }
\label{compRTfig}
\end{figure}

A comparison between this radiative transfer calculation and the forward model from a published atmospheric retrieval code \texttt{CHIMERA} \cp{Line13a} is shown in Figure \ref{compRTfig}. These two particular calculations use the same line lists, so this represents a test of just the radiative transfer and associated calculations. Note that the agreement is very good. \texttt{CHIMERA} calculates only absorption and emission, not scattering, so only cloud-free models can be directly compared. We have compared models that include clouds against previous similar calculations by \ct{Saumon08, Morley14a} and the agreement is also very good in regions where the line lists have not changed. Other tests comparing to other groups with different line lists and radiative transfer methods are beyond the scope of this work but would be important for understanding model uncertainties.

\end{document}